\documentclass[draftclsnofoot,onecolumn,12pt]{IEEEtran}
\ifCLASSINFOpdf
 \else
  \fi

\usepackage{cite} 
\usepackage{graphicx}
\usepackage[english]{babel}
\usepackage{amsmath}
\usepackage{amsthm}
\usepackage{algorithm,algpseudocode}

\usepackage{epsfig}
\usepackage{psfrag}
\usepackage{float}
\usepackage{enumerate}
\usepackage[table,xcdraw,dvipsnames]{xcolor}

\usepackage{times}
\usepackage{amssymb}
\usepackage{amsfonts}
\usepackage{bm}
\usepackage{amsmath}

\graphicspath{ {E:/Research_FSO/Paper_1/Latex_Journal/Fig/} }

\algrenewcommand\algorithmicforall{\textbf{foreach}}
\algrenewcommand\algorithmicindent{.8em}

\begin{document}
%
% paper title
% can use linebreaks \\ within to get better formatting as desired
%\title{Bare Demo of IEEEtran.cls for Journals}
%
%
% author names and IEEE memberships
% note positions of commas and nonbreaking spaces ( ~ ) LaTeX will not break
% a structure at a ~ so this keeps an author's name from being broken across
% two lines.
% use \thanks{} to gain access to the first footnote area
% a separate \thanks must be used for each paragraph as LaTeX2e's \thanks
% was not built to handle multiple paragraphs
%

\renewcommand{\textfraction}{0}
\title{A UAV-Mounted Free Space Optical Communication: Trajectory Optimization for Flight Time}
\author{Ju-Hyung Lee,~\IEEEmembership{Student Member,~IEEE,}
	Ki-Hong Park,~\IEEEmembership{Member,~IEEE,}
	Young-Chai Ko,~\IEEEmembership{Senior Member,~IEEE,}
	and~Mohamed-Slim Alouini~\IEEEmembership{Fellow,~IEEE}% <-this % stops a space
	
	\thanks{
	
	%This work was supported by the National Research Foundation of Korea (NRF) grant funded by the Korea government (MSIT)(NRF-2018R1A2B2007789).
	%This work was supported by King Abdullah University of Science and Technology.
	%This work is to be presented in part at the 2019 IEEE Globecom Workshops.
	
	J.-H. Lee, and Y.-C. Ko are with the School of Electrical and Computer Engineering, Korea University, Seoul, Korea (Email: leejuhyung@korea.ac.kr; koyc@korea.ac.kr)   
	
	K.-H. Park, and M.-S. Alouini are with the Electrical Engineering Program, Computer, Electrical, Mathematical Sciences and Engineering Division, King Abdullah University of Science and Technology (KAUST), Thuwal, Makkah Province, Kingdom of Saudi Arabia (Email: kihong.park@kaust.edu.sa; slim.alouini@kaust.edu.sa).
	
	%Part of this work is submitted to the IEEE International Conference on Communications, Shanghai, P.R. China, May. 2019.
	%}% <-this % stops a space
	
	%\thanks{Manuscript received April XX, 20XX; revised January XX, 20XX.}
}
}

\maketitle

\begin{abstract}
%\boldmath
In this work, we address the trajectory optimization of a fixed-wing unmanned aerial vehicle (UAV) using free space optical communication (FSOC). 
Here, we focus on maximizing the flight time of the UAV by considering practical constraints for wireless  UAV communication, including limited propulsion energy and required data rates.
We find optimized trajectories in various atmospheric environments (e.g., moderate-fog and heavy-fog conditions), while also considering the channel characteristics of FSOC.
In addition to maximizing the flight time, we consider the energy efficiency maximization and operation-time minimization problem to find the suboptimal solutions required to meet those constraints.
Furthermore, we introduce a low-complexity approach to the proposed framework.
In order to address the optimization problem, we conduct a bisection method and sequential programming and  introduce a new feasibility check algorithm. 
Although our design considers suboptimal solutions owing to the nonconvexity of the problems, our simulations indicate that the proposed scheme exhibits a gain of approximately 44.12\% in terms of service time when compared to the conventional scheme.
\end{abstract}
% IEEEtran.cls defaults to using nonbold math in the Abstract.
% This preserves the distinction between vectors and scalars. However,
% if the journal you are submitting to favors bold math in the abstract,
% then you can use LaTeX's standard command \boldmath at the very start
% of the abstract to achieve this. Many IEEE journals frown on math
% in the abstract anyway.

% Note that keywords are not normally used for peerreview papers.

\IEEEpeerreviewmaketitle

\section{Introduction}
Recently, unmanned aerial vehicles (UAVs) have attracted a great deal of attention in the area of wireless communication networks.
Owing to their mobility and flexibility, UAVs can be dispatched as a mobile entity in a network.
It can provide new opportunities in various communication applications and have the capability to complement conventional fixed networks.  
UAV-assisted communications can efficiently support already existing terrestrial communication infrastructure, including data offloading at a hot spot \cite{Intro_1, Intro_18}. 
For example, in case of a catastrophic event, UAV-mounted infrastructures can temporarily support service recovery initiatives and local interim communication facilities for potentially damaged infrastructures \cite{Intro_9}. 
In addition, UAV-assisted relaying can help to extend base station connectivity, from one station to another, in situations where the nodes are widely scattered and/or obstacles such as hills or large buildings are present \cite{Intro_2}.

Principally, we envision that most future networks will be highly user-centric, increasing the user demand information due to an exponential increase in the internet of things (IoT), fifth-generation (5G) mobile networks, and beyond-5G wireless networks. 
Accordingly, UAV-mounted fronthaul and backhaul frameworks have been considered as a promising approach to handle unexpected or temporarily large amounts of information that is commonly required by user terminals \cite{Intro_8, Intro_10}.
By leveraging the UAV-mounted fronthaul and backhaul links, large-scale projects are now underway in several research groups and information technology companies, including \textit{Global Access to the Internet for All (GAIA)} \cite{Intro_11}, \textit{Project Loon} by Google \cite{Intro_13}, and \textit{internet.org} by Facebook \cite{Intro_12,  Intro_6}.

%%%%%%%%%%%%%%%%%%%%%%%%%%%%%%%%%%%%%%%%%%%%%%%%%%%%%%%%%%%%%%%%%%%%%%%%%%%%%%%%%%%%%%%%%%%%%%%%%%%%%%%%%
\subsection{State-of-the-Art Literature} \label{Introduction_sub1}
%%%%%%%%%%%%%%%%%%%%%%%%%%%%%%%%%%%%%%%%%%%%%%%%%%%%%%%%%%%%%%%%%%%%%%%%%%%%%%%%%%%%%%%%%%%%%%%%%%%%%%%%%
By exploiting the high maneuverability of UAVs, current recent research activities have focused on optimizing mobile-UAV enabled wireless networks, such as placement or trajectory design, so that the UAV performs efficiently in accordance with several specific objectives \cite{Intro_3, Intro_4, Intro_5, Intro_20}.
User scheduling and association, UAV trajectory, and transmit power were all jointly optimized in \cite{Intro_3} to maximize the minimum average data rate among all users.
The authors of \cite{Intro_4} studied the possibility of minimizing mobile energy consumption over bit allocation for uplink, downlink, and computational processes as well as over the UAVs' trajectory design.
In \cite{Intro_5}, a framework was proposed to jointly optimize the three dimensional (3D) placement and mobility of UAVs, device-UAV association, and uplink power control in order to enable reliable uplink communications for IoT devices using a minimum total transmit power.
The optimal 3D trajectory of each UAV was investigated in such a way that the total energy consumed for the movement of the UAVs was minimized while retaining the support of the IoT devices.
%The UAV flight was thus optimized regarding those specific requirements.

In addition to the aforementioned scenarios, particularly where UAVs make use of RF communication, in other studies, free space optical communication (FSOC) mounted on UAVs has been discussed \cite{Intro_6, Intro_7, Intro_8}.
In 5G and future wireless network scenarios, optical wireless communications are currently gaining rapid  interest as an attractive alternative for providing a wide range of free spectrum bands to overcome the demands of the RF spectrum scarcity.
FSOC has several known advantages, such as an unlicensed broad spectrum, immunity to electromagnetic interference, and security. 
For example, by taking advantage of these, Facebook has launched project \textit{Aquila} in the hope of providing internet access to the world, by leveraging UAVs such that free space laser communication systems may be adopted \cite{Intro_13, Intro_6}.
Academic research groups have also suggested some specific platforms where UAVs can carry out various missions \cite{Intro_7,Intro_8}.
The authors of \cite{Intro_8} investigated the performance of FSOCs with UAVs subjected to different weather conditions and a broad range of system parameters. 
They were able to verify that FSOC-based vertical backhaul/fronthaul frameworks had the capacity to support transmission rates that were higher than the baseline alternatives, hence considered as an up-and-coming solution to satisfy the emerging backhaul/fronthaul requirements. 

There are, however, challenges in the wireless communications with a UAV, as it is difficult to recharge propulsion fuel or electric power during a UAV flight.\footnote{Electric power can be obtained through energy harvesting. Energy harvesting, however, falls outside the scope of this paper.} 
Flight time is also particularly limited owing to the finite energy capacity of UAVs, if energy-harvesting units are not utilized. 
Yet, this may not be a factor in applications requiring only a short running time, such as the duration of an entertainment event.
On the other hand, when long-term operations are considered (e.g., traffic monitoring, border surveillance, and environmental sensing), the UAV would need to fly as efficiently as possible in order to maximize its flight time.
Prior research \cite{Intro_14,Intro_15,Intro_16,Intro_19} has therefore been focused on the energy-saving issues of UAV communications in applications where long-term operations are needed.
The authors of \cite{Intro_14} concentrated on a UAV-enabled data collection system, where the UAV operates to collect a certain amount of information from the ground terminal at a fixed location. 
For ground-to-UAV wireless communications, the work showed that the transmission energy reduction of the ground terminal and the higher propulsion energy consumption of the UAV are in a trade-off relationship.
For a similar purpose, an optimal trade-off between the communication and the computational energy was attained for multiple UAVs \cite{Intro_15}.
This work showed that the system operation time extended significantly by minimizing the communication distance and the amount of data, at the expense of increased computational costs. 
In \cite{Intro_16}, UAVs were applied in a cognitive radio system to solve for the RF spectrum scarcity.
The authors considered the two main challenges of efficient energy management and opportunistic spectrum access and then proposed an energy-efficient solution by considering the hover and communication energies used.
The objective of the paper was to determine an optimized 3D location, where the UAV could transfer its data with minimum energy consumption, while avoiding any interference with the transmission of primary spectrum owners.

%%%%%%%%%%%%%%%%%%%%%%%%%%%%%%%%%%%%%%%%%%%%%%%%%%%%%%%%%%%%%%%%%%%%%%%%%%%%%%%%%%%%%%%%%%%%%%%%%%%%%%%%%
\subsection{Motivation \& Major Contributions} \label{Introduction_sub2}
%%%%%%%%%%%%%%%%%%%%%%%%%%%%%%%%%%%%%%%%%%%%%%%%%%%%%%%%%%%%%%%%%%%%%%%%%%%%%%%%%%%%%%%%%%%%%%%%%%%%%%%%%

\textcolor{Black}{Recently, the FSOC-based backhaul network has appealed a lot of attention in the literature. 
	In Facebook's \textit{Aquila} project \cite{Intro_6,Intro_27,Intro_28}, FSOC is employed at the source to UAV-assisted relay link and UAV-to-UAV link, to support the high throughput in the air-to-ground and air-to-air channel. 
	Especially in \cite{Intro_27,Intro_28}, Facebook has published a study on long endurance of UAV as well as energy-optimized trajectory planning for the project.
	However, quality of service (QoS) requirements to the air-to-ground communications are not considered in these works.}
Motivated by the aforementioned FSOC-based vertical backhaul framework \cite{Intro_8} and the Facebook's \textit{Aquila} project, we consider the FSOC-enabled backhaul network with the help of UAV, which can offer data rates higher than the baseline alternatives, and thus can be considered as a promising solution to the emerging wireless backhauling as discussed in \cite{Intro_22}. 
In our previous paper, we have discussed roughly the UAV-mounted FSOC system and solved the energy efficiency optimization problem in \cite{App_1}.
Especially focused on geometric loss on FSOC, we have shown the energy efficient trajectory of UAV.

\textcolor{Black}{
	In this paper, we deal with practical UAV-assisted backhaul networks\footnote{\textcolor{Black}{UAV-assisted backhaul networks have been dealt in Google's \textit{Loon} project \cite{Intro_26} as well as in report for future aerial communication networks \cite{Intro_23}.
			Furthermore, FSOC-based backhauling in wireless networks has been covered in various works \cite{Intro_25,Intro_24}.}} and solve optimization problems related to this application.
	In the UAV-assisted backhaul networks, the service time that meets a specific service requirement needs to be maximized.}
Since the energy consumption of UAV is a major challenge that limits its flight time, therefore, it is essential that further insight is gained into the average power consumption of UAV under different movement postures, such as hovering, moving, and circling \cite{Intro_8}.

\begin{figure}[t]
	\centering
	\includegraphics[width=3.4in]{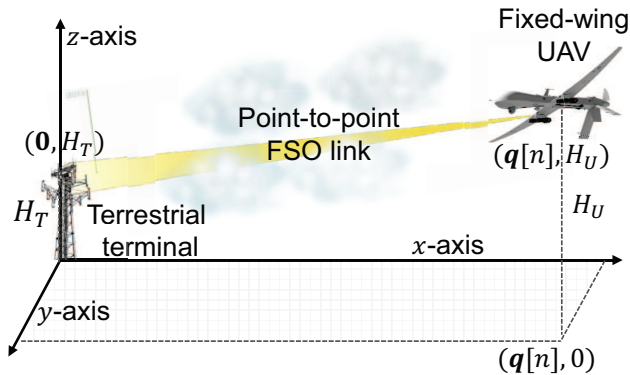}
	\caption{Point-to-point FSOC illustration between a UAV and a terrestrial terminal.}
	\label{Fig1}
	%    \vspace{-1.0em}
\end{figure}

In such a scenario, we focus on energy consumption constraints and aim to maximize flight-time (e.g., service time).
Furthermore, to utilize the mobility of UAV in a practical UAV-assisted backhaul network, the access procedure to the service boundary satisfying QoS requirements should be considered.
Accordingly, we consider not only the service time maximization in the service region, but also the entry/exit energy minimization of the UAV in this paper.
The main contributions of this work are summarized as follows:

\begin{enumerate}
	\item  We look into the scenario illustrated in Fig. \ref{Fig1}, where the UAV supports communications with the terrestrial terminal via point-to-point FSOC links.
	Specifically, we address flight time maximization under limited energy and FSOC data rate requirements to determine the optimized trajectory for which the UAVs can fly as long as possible while satisfying a required data rate.
	To the best of our knowledge, no other published literature has dealt with the FSOC-UAV trajectory problem except for our work \cite{App_1}.
	\item Based on the flight time maximization framework, we present energy efficiency maximization and operation-time minimization as an extension to other applications.
	\item We then introduce a low-complexity scheme based on rotational transformations to operate with less complexity despite the large number of time.  
	We also investigate the complexity analysis of both the proposed flight time maximization and the low-complexity scheme. 
	\item Under different atmospheric conditions (e.g., heavy-fog or moderate-fog conditions), we validate by carrying out simulations of the proposed schemes in order to determine their superiority over conventional schemes.
\end{enumerate}

%%%%%%%%%%%%%%%%%%%%%%%%%%%%%%%%%%%%%%%%%%%%%%%%%%%%%%%%%%%%%%%%%%%%%%%%%%%%%%%%%%%%%%%%%%%%%%%%%%%%%%%%%
\subsection{Outline} \label{Introduction_sub3}
%%%%%%%%%%%%%%%%%%%%%%%%%%%%%%%%%%%%%%%%%%%%%%%%%%%%%%%%%%%%%%%%%%%%%%%%%%%%%%%%%%%%%%%%%%%%%%%%%%%%%%%%%
The rest of this paper is organized as follows: 
Section \ref{system model} describes the system model, including the FSOC channel model and the rate model for the UAV using FSOC. 
In Section \ref{Body1}, we formulate the optimization problem by considering the data rate requirements and service range of flight time maximization.
Accordingly, we propose a suboptimal algorithm, by which we then extend the framework to other applications.
We also conduct a complexity analysis and, based on the results, introduce a low-complexity scheme for the framework in Section \ref{Body2}.
We then validate the quality of the proposed schemes using the numerical results in Section \ref{numerical result}. 
Finally, we conclude the paper in Section \ref{conclusion}. 

% needed in second column of first page if using \IEEEpubid
%\IEEEpubidadjcol

\textit{Notation:}
Throughout this paper, we use the normal-face font to denote scalars, and boldface font to denote vectors.
We use $\mathbb{R}^{D\times 1}$ to represent the $D$-dimensional space of real-valued vectors.
We also use $\|\cdot\|$ to denote the $L^2$-norm (i.e., an Euclidean norm) and $\mathrm{log}(\cdot)$ to represent the natural logarithm.
The expression $O(\cdot)$ stands for describing the Big O notation.
The function $\mathrm{floor}\{x\}$ is the greatest integer function that takes a real number $x$ as an input, which generates an output that is less than or equal to $x$.

\section{System Model}  \label{system model}

%%%%%%%%%%%%%%%%%%%%%%%%%%%%%%%%%%%%%%%%%%%%%%%%%%%%%%%%%%%%%%%%%%%%%%%%%%%%%%%%%%%%%%%%%%%%%%%%%%%%%%%%%
\subsection{UAV System Model}
%%%%%%%%%%%%%%%%%%%%%%%%%%%%%%%%%%%%%%%%%%%%%%%%%%%%%%%%%%%%%%%%%%%%%%%%%%%%%%%%%%%%%%%%%%%%%%%%%%%%%%%%%

Fig. 1 shows the air-to-ground FSOC system, which enables communication between the terrestrial terminal and the UAV. 
This point-to-point FSOC communication can be understood as either ground-to-UAV communication or UAV-to-ground communication.
Note that the UAV can serve as a aerial base station in FSOC-enabled backhaul networks.
Considering the UAV's mobility, we set up the problem in 3D Cartesian coordinates.
\textcolor{Black}{Here, we assume that the terrestrial terminal is located at the origin in the $x$-$y$ plane at a fixed altitude $H_{T}$ on the $z$-axis, and the UAV flies horizontally in the $x$-$y$ plane at a constant altitude $H_{U}$ on the $z$-axis.}\footnote{Although the values of the z-axis can be changed depending on which UAV platform is operated, for simplicity, we assume a low-altitude platform and a fixed altitude.}
For ease of optimization, we deal with the system as a discrete-time model\footnote{Based on time step size $\delta_{t}$ and time $T$ (or $t$), time slot $N$ (or $n$) can be determined according to $T=\delta_{t} \cdot N$ (or $t=\delta_{t} \cdot n$).}, similarly to \cite{Intro_2,System_1}. 
\textcolor{Black}{With $\delta_t$ as a certain time step, the UAV position vector can be expressed as $\mathbf{q}[n]=\left( x[n], y[n] \right)^{T}\in\mathbb{R}^{2\times 1} $ located at a constant altitude $H_{U}$ on the $z$-axis in a time slot $n=0,\cdots, N+1$.
	The distance between the terrestrial terminal and the UAV can be represented by $d[n]=\sqrt{(H_{U}-H_{T})^{2}+\Vert\mathbf{q}[n]\Vert^{2}}$.}

We denote the UAV velocity vector as $\mathbf{v}[n]$ and UAV acceleration vector as $\mathbf{a}[n]$.
Accordigly, we formulate the discrete UAV state using the Taylor approximation, as described in \cite{System_1}, 

\begin{eqnarray}
\mathbf{v}[n+1] \!\!\!\!&=&\!\!\!\! \mathbf{v}[n]+\mathbf{a}[n]\delta_t, \ n=0,\cdots, N, \label{C_v&a} \\
\mathbf{q}[n+1] \!\!\!\!&=&\!\!\!\! \mathbf{q}[n]+\mathbf{v}[n]\delta_t + \frac{1}{2}\mathbf{a}[n]\delta_t^2, \ n=0,\cdots, N.  \label{C_q&v&a}
\end{eqnarray}	

Let us denote $\mathbf{q}_{\mathrm{I}}$ and $\mathbf{q}_{\mathrm{F}}$ as the initial and final positions and $\mathbf{v}_{\mathrm{I}}$ and $\mathbf{v}_{\mathrm{F}}$ as the initial and final velocities of the UAV, which can be expressed as

\begin{eqnarray}
\mathbf{q}[0]\!\!&\!\!=\!\!&\!\!\mathbf{q}_{\mathrm{I}},\ \mathbf{q}[N+1]=\mathbf{q}_{\mathrm{F}}, \label{C_qI&qF} \\
\mathbf{v}[0]\!\!&\!\!=\!\!&\!\!\mathbf{v}_{\mathrm{I}},\ \mathbf{v}[N+1]=\mathbf{v}_{\mathrm{F}}. \label{C_vI&vF} 
\end{eqnarray}	
Note that $n=0$ and $n=N+1$ denote the initial time slot and final time slot, respectively.

Moreover, depending on the UAV's characteristics, the speed and acceleration of the UAV staying aloft in still air are limited, mathematically, to

\begin{eqnarray}
\!\!&\!\! \|\mathbf{v}[n]\| \leq V_{\max},   \!\!&\!\! \|\mathbf{a}[n]\|  \leq A_{\max}, \ \forall n, \label{C_Vmax&Amax} \\ %\label{C_Amax}
\!\!&\!\! V_{\min}  \leq  \|\mathbf{v}[n]\|, \!\!&\!\! \ \forall n, \label{C_Vmin} %\label{C_Vmin&Vmax} 
\end{eqnarray}
where $V_{\max}$ and $A_{\max}$ are the maximum velocity and maximum acceleration, respectively, and $V_{\min}$ indicates the minimum velocity.

\begin{figure*}
	%\vspace{-1.5em}
	\begin{equation}
	E_T=\sum_{n=1}^{N} \left[ c_1\|\mathbf{v}[n]\|^3 + \frac{c_2}{\|\mathbf{v}[n]\|}\left(1+\frac{\|\mathbf{a}[n]\|^2-\frac{(\mathbf{a}^T[n]\mathbf{v}[n])^2}{\|\mathbf{v}[n]\|^2}}{g^2}\right) \right] \cdot \delta_t + \frac{m}{2} \cdot \left( \|\mathbf{v}[N+1]\|^2 - \|\mathbf{v}[0]\|^2 \right) \ [\mathrm{Joule}], 
	\label{Energy}
	\end{equation} \hrule	
	%\vspace{-1.2em}
\end{figure*}

In the UAV system, the energy consumption model is key to optimizing the flight time. 
We consider the propulsion energy for a flight as $E_f$ and the communication energy for a signal processing as $E_c$.
Note that, in the energy consumption model, we assume $E_c$ to be a constant, as it is known to be much smaller than $E_f$ in practical scenarios (i.e., $E_f \gg E_c$) \cite{System_7,System_1} . 
Therefore, we simply use total energy as $E_T \simeq E_f$.

Following \cite{System_1,System_8}, the discrete-time energy consumption model for a fixed-wing UAV can be written as \eqref{Energy} at the top of next page. 
Note that $c_1$ and $c_2$ are two parameters related to the effect of aircraft weight, its wing area, and the air density.
Here, $g$ is the acceleration due to gravity (9.8 m/s$^{2}$) and $m$ is the mass of the UAV. 
\textcolor{Black}{Note that $E_T$ in \eqref{Energy} can be considered as energy in Joule.}
%\footnote{\textcolor{Black}{In this paper, we assume the time step size $\delta_t = 1$ [s]. Therefore, we can consider $E_T$ as the consumption energy model of UAV.}}
The upper bound of the energy consumption model is represented by $\sum_{n=1}^{N} c_1\cdot\delta_t\|\mathbf{v}[n]\|^3 \!+\! \frac{c_2\cdot\delta_t}{\|\mathbf{v}[n]\|} \left(\! 1\!+\!\frac{\|\mathbf{a}[n]\|^2}{g^2} \!\right) + K_{E}$ assuming with $\mathbf{a}^T[n]\mathbf{v}[n]=0$.
Note that the kinetic energy of the UAV, $K_{E}=\frac{m}{2} \cdot \left( \|\mathbf{v}[N+1]\|^2 - \|\mathbf{v}[0]\|^2 \right)$, that is the last term in \eqref{Energy}, will be set to zero if we assume that the initial velocity and final velocity are the same.

%%%%%%%%%%%%%%%%%%%%%%%%%%%%%%%%%%%%%%%%%%%%%%%%%%%%%%%%%%%%%%%%%%%%%%%%%%%%%%%%%%%%%%%%%%%%%%%%%%%%%%%%%
\subsection{FSOC System Model}
%%%%%%%%%%%%%%%%%%%%%%%%%%%%%%%%%%%%%%%%%%%%%%%%%%%%%%%%%%%%%%%%%%%%%%%%%%%%%%%%%%%%%%%%%%%%%%%%%%%%%%%%%

By understanding the channel characteristics for FSOC, we carefully set up the FSOC system. 
Here, we assume a line of sight (LoS) between the UAV and the terrestrial terminals\footnote{In the air-to-ground channel, the LoS is easily obtainable at a given altitude where the UAV is flying.} such that multipath propagation is not considered. 
We also assume that a fast fading that may occur with the location and movement of UAVs is adequately compensated.
Therefore, we base the time-varying nature of the system solely on the atmospheric loss model of FSOC \cite{System_2}. 

\subsubsection{Attenuation of the FSOC Channel}
Atmospheric losses of optical signal propagation are determined by environmental conditions such as absorption or scattering effects. 
According to the Beer-Lambert Law, the signal attenuation obtained owing to weather conditions\footnote{\textcolor{Black}{Out of various environmental factors, atmospheric attenuation is typically dominated by fog, as the particle size of fog is comparable to the wavelength of interest in FSOC \cite{System_2, System_5}. Thus, we mainly focus on the effects of fog for signal attenuation based on Beer-Lambert Law. Note that if other attenuation factors, e.g., rain, snow and haze, need to be considered, the optimization framework can be solved by adjusting only the parameters, e.g., $\beta$ or $k_2$.
		%Note also that, unlike FSOC, RF communications (especially, mm-wave communication) are not attenuated by fog, however, are highly attenuated by water molecules such as rainfall \cite{System_9}.
}} can be expressed as \cite{System_2}

\begin{equation}
\tilde{\beta}_{\mathrm{FSO}}=\frac{3.91}{V}\left(\frac{\lambda}{550 \ \mathrm{nm}}\right)^{-p} \  \mathrm{[dB/km]}, 
\end{equation}
where $\lambda$ is the wavelength (set to 1550 nm in this paper) and $V$ is the visibility in kilometers.
The size distribution coefficient of the scattering $p$ is derived from the Kruse model \cite{System_6}. 
Based on Beer's law, the atmospheric loss is given by \cite{System_10, System_5} 

\begin{eqnarray}
h_{\mathrm{FSO}}=e^{-\beta_{\mathrm{FSO}} \cdot d[n]}, \label{Path loss} 
\end{eqnarray}
where $\beta_{\mathrm{FSO}} = \frac{\tilde{\beta}_{\mathrm{FSO}} \log{10}}{10^4}\ \mathrm{[m^{-1}]}$. 

\subsubsection{Rate Model for FSOC}
To the best of our knowledge, the capacity of FSOCs has yet to be investigated in closed form, although the bound for the capacity of FSOCs is currently being studied in several papers \cite{System_3, System_4}.
Here, the lower bound (introduced in \cite{System_3}) is used as the transmission rate model between the UAV and the terrestrial terminal.
In FSOC, the average optical signal-to-noise ratio (ASNR) is represented as $\gamma=\frac{\varepsilon}{\sigma}$, where the average-to-peak ratio (APR) is $\alpha=\frac{\varepsilon}{\Lambda}$, $\sigma$ is the noise variance, $\Lambda$ is the peak optical power, and $\varepsilon$ is the average optical power \cite{System_3}. 

When ASNR $\overline{\gamma}=h_{\mathrm{FSO}}\cdot\gamma$ is obtained, we can express the discrete-time rate model for FSOC in [bps] as

\begin{eqnarray}
\!\!\!\!&\!\!\!\! \!\!&\!\! R_{\mathrm{FSO}}(\mathbf{q}[n])=\frac{B_{\mathrm{FSO}}}{2 \mathrm{log}2} \log\left( 1+k_1 e^{-k_2\sqrt{(H_{U}-H_{T})^2+\|\mathbf{q}[n]\|^2}}\right),   \nonumber \\
\!\!\!\!&\!\!\!\! \!\!&\!\!   \label{Rate}
\end{eqnarray}
where $B_{\mathrm{FSO}}$ is the bandwidth of FSOC, and $k_1$ and $k_2$ can then be expressed as 

\begin{eqnarray}
k_1 \!=\! \begin{cases}
\frac{e^{2\alpha\mu^{*}}}{2\pi e}\left( \frac{1-e^{-\mu^{*}} }{\mu^{*}}\right)^2 \frac{\gamma^2}{\alpha^2},  
\!\!\!&\!\!\! (0\!<\!\alpha\!<\!\frac{1}{2}) \\
\frac{\gamma^2}{2\pi e \alpha^2}, 
\!\!\!&\!\!\! (\frac{1}{2}\!<\!\alpha\!<\!1)
\end{cases} 
,\!\!&\!\! \mbox{and}\:\: k_2=2 \cdot \beta_{\mathrm{FSO}} , & \nonumber \label{k1,k2} 
\end{eqnarray}
where $\alpha=\frac{1}{\mu^{*}}-\frac{e^{- \mu^{*}}}{1-e^{-\mu^{*}}}$ is a free parameter and $\mu^{*}$ is a solution to the previous equation. 
As described in \cite{System_3}, for predetermined value $\alpha$, the uniqueness
and existence of $\mu^{*}$ can be shown easily. 
Note that $\mu^{*}$ can be found with root-finding algorithms, e.g., bisection method and interpolation \cite{System_11}.

%%%%%%%%%%%%%%%%%%%%%%%%%%%%%%%%%%%%%%%%%%%%%%%%%%%%%%%%%%%%%%%%%%%%%%%%%%%%%%%%%%%%%%%%%%%%%%%%%%%%%%%%%%%%%%%%%%%%%%%%%%%%%%%%%%%%%%%%%%%%%%%%%%%%%%%%%%%%%%%%%%%%%%%%%%%%%%%%%%
\section{Flight Time Maximization} \label{Body1}

\begin{figure}[t]
	\centering
	\includegraphics[width=3.4in]{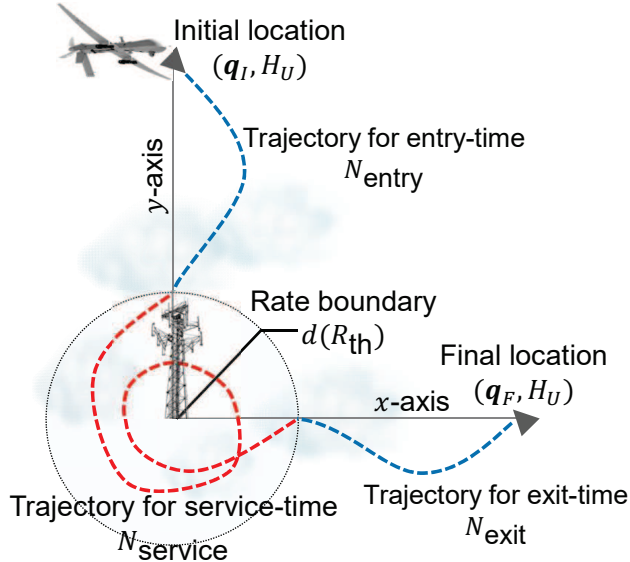}
	\caption{Flight time maximization diagram.}
	\label{fig:visio2}
\end{figure}

This section introduces the optimization framework for maximizing the flight time of FSOC-UAV considering a data rate requirements illustrated in Fig. \ref{fig:visio2}.
Here, we discuss two optimization problems for the service time maximization and the entry/exit energy minimization, and suggest a new feasibility algorithm for troubleshooting this nonconvex problem.

Note that the service time ($N_{\mathrm{service}}$) is the time when the UAV transmits data at a rate that is higher than the required minimum data rate ($R_{\mathrm{th}}$) to the ground terminal. 
We can, therefore, establish a maximum flight radius $d(R_{\mathrm{th}})$ for the service boundary, such that

\begin{equation}
R_{\mathrm{FSO}}(\mathbf{q}[n]) \geq R_{\mathrm{th}}, \hspace{0.05in} n \in N_{\mathrm{service}}. 
\end{equation}
Therefore, we can derive the service boundary using \eqref{Rate} as 

\begin{equation}
\begin{split}
d(R_{\mathrm{th}}) = \ &
\Bigg( \dfrac{1}{(k_2)^2} \cdot \bigg(\mathrm{log} \Big( \frac{1}{k_1}e^{\frac{2\mathrm{log}2 \cdot R_{\mathrm{th}}}{B_{\mathrm{FSO}}}}-\frac{1}{k_1} \Big) \bigg)^2 \\
&- (H_{U}-H_{T})^2 \Bigg)^{1/2}. 
\end{split}
\end{equation}

%\begin{eqnarray}
%\!\!\!&\!\!\! \!\!\!&\!\!\! \|\mathbf{q}[n]\| \leq d(R_{\mathrm{th}}) \!=\!
%\sqrt{ \dfrac{1}{(k_2)^2}\left(\mathrm{log}\!\left(\frac{1}{k_1}e^{\frac{2\mathrm{log}2 R_{\mathrm{th}}}{B_{\mathrm{FSO}}}}-\frac{1}{k_1} \right)\right)^2 \!\!-\! H^2}, \nonumber\\
%\!\!\!&\!\!\! \!\!\!&\!\!\! \hspace{0.05in} n \in N_{service} \nonumber
%\end{eqnarray}

In order to consider an actual flight situation, we assume that the UAV flies at a given total energy ($E_{\mathrm{total}}$).
Note that we adopt the following notations to better understand the continuous variables in the optimization problems: the position of UAV $\mathcal{Q} = \{\mathbf{q}[n], \ \forall n\}$, the velocity of UAV $\mathcal{V} = \{\mathbf{v}[n], \ \forall n\}$, and the acceleration of UAV $\mathcal{A} = \{\mathbf{a}[n], \ \forall n\}$.
Accordingly, the flight time maximization problem can be formulated as follows: 

\begin{subequations}
	\begin{eqnarray}
	\!\!\!&\!\!\! \displaystyle \max_{ \scriptsize \mathcal{Q}, \mathcal{V}, \mathcal{A} } \!\!&\!\! N_{\mathrm{service}} 	\label{P0}\\	
	\!\!\!&\!\!\! \textrm{s.t} \!\!&\!\! \eqref{C_v&a}-\eqref{C_Vmin}, \label{C_P0} \nonumber\\ 
	\!\!\!&\!\!\! \!\!&\!\! \sum_{N_{\mathrm{total}}} \!\! c_1\|\mathbf{v}[n]\|^3 \!+\! \frac{c_2}{\|\mathbf{v}[n]\|} \left(\! 1\!+\!\frac{\|\mathbf{a}[n]\|^2}{g^2} \!\right) \!\leq\! \frac{E_{\mathrm{total}}}{\delta_t}, \nonumber\\
	\!\!\!&\!\!\! \!\!&\!\! \label{C1_P0}\\
	\!\!\!&\!\!\! \!\!&\!\! \| \mathbf{q}[n] \| \leq d(R_{\mathrm{th}}), \label{C2_P0}\\
	\!\!\!&\!\!\! \!\!&\!\! (n \!=\! N_{\mathrm{entry}}+1,\cdots,N_{\mathrm{total}}-N_{\mathrm{exit}}).  \nonumber
	\end{eqnarray}	
\end{subequations}

As shown in Fig. \ref{fig:visio2}, we use mainly two algorithms for this problem. 
First, we develop an algorithm that minimizes the energy consumption during the entry time ($N_{\mathrm{entry}}$) and exit time ($N_{\mathrm{exit}}$) such that higher energy consumption can be concentrated over the service area.
Second, we propose an algorithm that maximizes the service time ($N_{\mathrm{service}}$) when the UAV flies within the service range, using only its remaining energy. 
By utilizing these two algorithms, we optimize the service time in the total flight time $N_{\mathrm{total}}=N_{\mathrm{entry}}+N_{\mathrm{service}}+N_{\mathrm{exit}}$, for the given energy of a UAV. 
With these algorithms, the UAV can efficiently enter the target service region, and provide service to a terrestrial terminal while meeting the data rate requirements as long as possible. 

%%%%%%%%%%%%%%%%%%%%%%%%%%%%%%%%%%%%%%%%%%%%%%%%%%%%%%%%%%%%%%%%%%%%%%%%%%%%%%%%%%%%%%%%%%%%%%%%%%%%%%%%%
\subsection{Energy Minimization during Entry to the Target Boundary} \label{Body1_sub1}
%%%%%%%%%%%%%%%%%%%%%%%%%%%%%%%%%%%%%%%%%%%%%%%%%%%%%%%%%%%%%%%%%%%%%%%%%%%%%%%%%%%%%%%%%%%%%%%%%%%%%%%%%
\textcolor{Black}{
	We note that energy minimization problem is equivalent to power minimization problem with the fixed $\delta_t$. From now on, we omit $\delta_t$ in the objective function of energy minimization problem for simplicity.}
Under the target rate boundary with a radius of $d(R_{\mathrm{th}})$, the energy minimization during entry time can be expressed as follows:

\begin{subequations}
	\begin{eqnarray}
	\!\!\!\!\!\!\!&\!\!\!\!\!\! \displaystyle \min_{ \scriptsize \begin{array}{c}
		\mathcal{Q}, \mathcal{V}, \mathcal{A} 
		\end{array} } \!\!&\!\! P(\mathbf{v}[n],\|\mathbf{v}[n]\|,\mathbf{a}[n]) \label{P1} \\
	%\sum_{N_{\mathrm{entry}}} c_1\cdot\delta_t\|\mathbf{v}[n]\|^3 + \frac{c_2\cdot\delta_t}{\|\mathbf{v}[n]\|} \left( 1+\frac{\|\mathbf{a}[n]\|^2}{g^2} \right)	\nonumber \\
	%\!\!\!\!\!\!\!&\!\!\!\!\!\! \!\!&\!\! \label{P1} \\ 	
	\!\!\!\!\!\!\!&\!\!\!\!\!\! \textrm{s.t} \!\!&\!\! \eqref{C_v&a}-\eqref{C_Vmin}, \nonumber \\
	\!\!\!\!\!\!\!&\!\!\!\!\!\! \!\!&\!\! \| \mathbf{q}[N_{\mathrm{entry}}] \| \leq d(R_{\mathrm{th}}), \label{C_1_P1} \\
	\!\!\!\!\!\!\!&\!\!\!\!\!\! \!\!&\!\! (n=1,\ldots, N_{\mathrm{entry}}). \nonumber
	\end{eqnarray}	
\end{subequations}
Note that the constraints in \eqref{C_v&a}-\eqref{C_Vmax&Amax} and  \eqref{C_1_P1} are convex. 
Conversely, the objective in \eqref{P1} and the minimum velocity constraint \eqref{C_Vmin} are nonconvex. 
We therefore approximate the nonconvex portions to be convex introducing a slack variable set $\mathcal{\tau} = \{ \mathbf{\tau}[n], \forall n \}$, and using the first Taylor approximation \cite{System_1}. 
Using a slack variable and the given $N_{\mathrm{entry}}$, we can rewrite \eqref{P1} as follows:

\begin{subequations}
	\begin{eqnarray}	
	\!\!\!\!&\!\!\!\! \displaystyle \min_{ \scriptsize \begin{array}{c}
		\mathcal{Q},\mathcal{V}\\
		\mathcal{A},\mathcal{\tau}
		\end{array} } \!&\! \sum_{N_{\mathrm{entry}}} P(\mathbf{v}[n],\tau[n],\mathbf{a}[n]) \label{P1_Mod} \\
	\!\!\!\!&\!\!\!\! \textrm{s.t} \!&\! \eqref{C_v&a}-\eqref{C_Vmax&Amax}, \eqref{C_1_P1}, \nonumber \\
	\!\!\!\!&\!\!\!\! \!&\! V_{\min}\leq \mathbf{\tau}[n], \ \mathbf{\tau}[n]^{2} \leq \phi(\mathbf{v}[n]), \label{C1_P1_Mod} \\
	\!\!\!\!&\!\!\!\!   \!&\! (n=1,\ldots, N_{\mathrm{entry}}), \nonumber
	\end{eqnarray}
\end{subequations}
\textcolor{Black}{where $P(\mathbf{v}[n],\tau[n],\mathbf{a}[n]) \triangleq c_1\|\mathbf{v}[n]\|^3 + \frac{c_2}{\tau[n]}\left(1+\frac{\|\mathbf{a}[n]\|^2}{g^2}\right)$ is defined as the power model of UAV, $\tau[n]$ is the slack variable for troubleshooting the nonconvex portions of \eqref{C_Vmin}, and $\phi(\mathbf{v}[n])\!\!=\!\!\|\mathbf{v}_j[n]\|^2 + 2\mathbf{v}_j^T[n](\mathbf{v}[n]\!-\!\mathbf{v}_j[n])$ is the first Taylor approximation of $\|\mathbf{v}[n]\|^2$ at $\mathbf{v}_j[n]$. 
}
%Since $P(\mathbf{v}[n],\|\mathbf{v}[n]\|,\mathbf{a}[n]) \cdot \delta_t$ and $P(\mathbf{v}[n],\tau[n],\mathbf{a}[n]) \cdot \delta_t$ are equivalent to the upper bound of energy consumption model, $P(\mathbf{v}[n],\|\mathbf{v}[n]\|,\mathbf{a}[n])$ and $P(\mathbf{v}[n],\tau[n],\mathbf{a}[n])$ can be considered as the objective function of energy minimization problem.
%For convenience, we consider $P(\mathbf{v}[n],\|\mathbf{v}[n]\|,\mathbf{a}[n])$ and $P(\mathbf{v}[n],\tau[n],\mathbf{a}[n])$ as the objective function in the energy minimization problems.

For a given $\mathbf{v}_j[n]$, the convex optimization problem \eqref{P1_Mod} is the quadratically constrained quadratic program (QCQP).
We can solve the QCQP within a polynomial complex for a given $N_{\mathrm{entry}}$ using standard convex optimization solvers such as CVX \cite{Body_6}.
We can also solve \eqref{P1} via the sequential convex optimization of \eqref{P1_Mod} through iteratively updating the local point $\{ \mathbf{q}_j[n], \mathbf{v}_j[n] \}$ as in \cite{System_1, Body_4, Body_5}.
Note that the sequential convex optimization method has been proven to converge to at least one local optimal point \cite{System_1}.

In order to find the optimum $N_{\mathrm{entry}}$ for energy minimization during entry time, we use a bisection method by checking both the gradient of the energy (within the given constraints) and $N_{\mathrm{entry}}$ in each iteration.
We then repeatedly update $N_{\mathrm{entry}}$ until convergence criteria are satisfied. 
Algorithm \ref{Algorithm0} describes the detailed process of entry energy minimization.
Note that we can minimize the exit energy consumption during exit time, $N_{\mathrm{exit}}$, with the equivalent scheme above.

\begin{algorithm}[]
	\caption{Entry energy minimization.} \label{Algorithm0}
	\begin{algorithmic}[1]
		\State Initialize $N_{\mathrm{entry}}$, upper/lower bound $\overline{N}_{\mathrm{entry}}$/$\underline{N}_{\mathrm{entry}}$. 
		
		\While {$\overline{N}_{\mathrm{entry}}-\underline{N}_{\mathrm{entry}}>\epsilon$$^\dag$} 
		%$
		\State Solve \eqref{P1} with $(N_{\mathrm{entry}}+1)$ and $(N_{\mathrm{entry}}-1)$. 
		%\vspace{0.025in}
		\State Find the gradient $K \triangleq E_{\mathrm{entry+1}}-E_{\mathrm{entry-1}}$.$^\ddagger$ 
		%\vspace{0.025in}
		\If {$K \leq 0$ \\ \ \ \ \ } {$\underline{N}_{\mathrm{entry}}=N_{\mathrm{entry}}$.} 
		%\vspace{0.025in}
		\Else \\ \ \ \ \ \ \textbf{then} $\overline{N}_{\mathrm{entry}}=N_{\mathrm{entry}}$.
		\EndIf %\vspace{0.025in}
		\State Update $N_{\mathrm{entry}}=\dfrac{\overline{N}_{\mathrm{entry}}+ \underline{N}_{\mathrm{entry}}}{2}$. 
		\EndWhile
		\State \textbf{end} %\vspace{0.05in}
		\State Find the minimum value by comparing $E_{\mathrm{entry-1}}$, $E_{\mathrm{entry}}$, and $E_{\mathrm{entry+1}}$, and then determine the optimum value $N_{\mathrm{entry}}$.
		%\State Using the optimum value $N_{\mathrm{entry}}$, find the minimum value among  $E_{\mathrm{entry-1}}$, $E_{\mathrm{entry}}$, $E_{\mathrm{entry+1}}$. %\vspace{0.05in}
	\end{algorithmic}
	\hrule%\vspace{0.05in}
	$^\dag$ \footnotesize $\epsilon$ is the tolerance value for algorithm's termination point. \\ 
	$^\ddag$ \footnotesize $E_{\mathrm{entry-1}}$, $E_{\mathrm{entry}}$, and $E_{\mathrm{entry+1}}$ are the consumption energy of UAV during the time steps $N_{\mathrm{entry}}-1$, $N_{\mathrm{entry}}$, and $N_{\mathrm{entry}}+1$, respectively. 
\end{algorithm}

%%%%%%%%%%%%%%%%%%%%%%%%%%%%%%%%%%%%%%%%%%%%%%%%%%%%%%%%%%%%%%%%%%%%%%%%%%%%%%%%%%%%%%%%%%%%%%%%%%%%%%%%%
\subsection{Maximizing Service Time with Rate Requirements} \label{Body1_sub2}
%%%%%%%%%%%%%%%%%%%%%%%%%%%%%%%%%%%%%%%%%%%%%%%%%%%%%%%%%%%%%%%%%%%%%%%%%%%%%%%%%%%%%%%%%%%%%%%%%%%%%%%%%

%In order to optimize the service time by considering the rate requirement, we set up energy consumption as an objective function, for a fixed $N_{\mathrm{service}}$. 
%We first manage energy minimization within the data rate constraint and then find the service time to satisfy the total power constraint.
In order to optimize the service-time considering the rate requirement, we set up the energy consumption model as an objective function, instead of $N_{\mathrm{service}}$. 
We can intuitively understand that the flight time of the UAV can be maximized when flying with minimum energy.
Therefore, the problem can be formulated as follows:

\begin{subequations}
	\begin{eqnarray}
	\!\!\!\!\!\!\!&\!\!\!\!\!\!	 \displaystyle \min_{ \scriptsize \begin{array}{c}
		\mathcal{Q},\mathcal{V}, \mathcal{A}
		\end{array} } \!&\!
	\sum_{N_{\mathrm{service}}} P(\mathbf{v}[n],\|\mathbf{v}[n]\|,\mathbf{a}[n]) \label{P2} \\
	\!\!\!\!\!\!\!&\!\!\!\!\!\! \textrm{s.t} \!&\! \eqref{C_v&a}-\eqref{C_Vmin}, \nonumber \\
	\!\!\!\!\!\!\!&\!\!\!\!\!\! \!&\! \| \mathbf{q}[n] \| \leq d(R_{\mathrm{th}}),  \label{C1_P2}\\ 
	\!\!\!\!\!\!\!&\!\!\!\!\!\! \!&\! (n=1,\cdots,N_{\mathrm{service}}). \nonumber  
	\end{eqnarray}
\end{subequations}	
Note that $\mathbf{q}_{\mathrm{F(entry)}}$ and $\mathbf{v}_{\mathrm{F(entry)}}$ are the final position and velocity, respectively, for the entry energy minimization scheme and $\mathbf{q}_{\mathrm{I(exit)}}$ and $\mathbf{v}_{\mathrm{I(exit)}}$ are the initial position and initial velocity, respectively, for the exit energy minimization. 
These values are used as the initial and final positions and velocities in \eqref{P2}.

\textcolor{Black}{
	Given that the rate boundary constraint \eqref{C1_P2} is convex, we apply Taylor approximation, introduce slack variable, and then solve \eqref{P2} with sequential convex programming, in the equivalent method for \eqref{P1_Mod}.
	In order to solve the problem of \eqref{P2} with the sequential programming, initial values of $\mathbf{v}_{\mathrm{service}}[n], \forall n$ which belong to the feasible set of \eqref{P2}, are required.
	However, it has been challenging to intuitively set the initial value due to the constraint of \eqref{C1_P2}.
	To find the initial values ($\mathbf{v}^{\star}_{\mathrm{service}}[n], \forall n$) in the feasible set, proper feasibility check is required before applying the sequential programming to \eqref{P2}.
	Note that the constraints in \eqref{C_Vmin} are nonconvex and thus it is challenging to use conventional feasibility check in this problem \cite{Body_1}.
	We therefore propose Algorithm \ref{Algorithm1} for the nonconvex feasibility check and use the algorithm for finding initial values to solve \eqref{P2}.}

% (algorithm for non-convex feasibility check)    
%\begin{algorithm}[]
%\caption{Feasibility check by minimizing norm distance between two sets.} \label{Algorithm1}
%\begin{algorithmic}[1]
%\State Initialize tolerance $\epsilon$ 
%\While{\textcolor{black}{
%%the partial decrease for the objective value is above a tolerance}}
%\State Solve the problem as follow 
%\begin{eqnarray}
%%	\!\!&\!\! \textrm{s.t} & \mathbf{x}_1[n] \in \mathcal{D}^k, \ \mathbf{x}_2[n] \in \mathcal{P}^k, \ \forall n, \nonumber 
%\end{eqnarray}	
%where $\mathcal{D}^k$ and $\mathcal{P}^k$ are the feasible sets satisfying the constraints of $\mathbf{x}_1[n]$ and $\mathbf{x}_2[n]$, $\forall n$, respectively.
%\If{optimal value $\leq$ $\epsilon^{\dag}$}{\ feasible} 
% \State \textcolor{black}{\textbf{break};}
% \Else \ {infeasible} 
% % \State \textcolor{black}{Update the variables based on both the objective function and the constraints in the problem}
% \State \textcolor{black}{Update $\mathcal{D}^{k+1}$ and $\mathcal{P}^{k+1}$ based on the result of step2, and the iteration number $k=k+1$, }
% \EndIf
% \EndWhile
% %\State \textbf{end} 
% \end{algorithmic}
% \hrule
% $^\dag$ \footnotesize $\epsilon$ is the convergence threshold of the algorithm.
% \end{algorithm}   

\begin{algorithm}[]
	\caption{Feasibility check by minimizing norm distance between two sets.} \label{Algorithm1}
	\begin{algorithmic}[1]
		%\State Initialize tolerance $\epsilon$ 
		\State Solve the problem as follows 
		\begin{eqnarray}
		\!\!&\!\! \displaystyle \min & \sum (\| \mathbf{x}_1[n] - \mathbf{x}_2[n] \|^2) \nonumber \\ 	
		\!\!&\!\! \textrm{s.t} & \mathbf{x}_1[n] \in \mathcal{D}, \ \mathbf{x}_2[n] \in \mathcal{P}, \ \forall n, \nonumber 
		\end{eqnarray}	
		where $\mathcal{D}$ and $\mathcal{P}$ are the feasible sets satisfying the constraints of $\mathbf{x}_1[n]$ and $\mathbf{x}_2[n]$, $\forall n$, respectively.
		\If{optimal value $\leq$ $\epsilon^{\dag}$}{\ feasible} 
		%\State \textcolor{black}{\textbf{break};}
		\Else \ {infeasible} 
		% \State \textcolor{black}{Update the variables based on both the objective function and the constraints in the problem}
		%\State \textcolor{black}{Update $\mathcal{D}^{k+1}$ and $\mathcal{P}^{k+1}$ based on the result of step2, and the iteration number $k=k+1$, }
		\EndIf
		%\EndWhile
		%\State \textbf{end} 
	\end{algorithmic}
	\hrule
	$^\dag$ \footnotesize $\epsilon$ is the convergence threshold of the algorithm.
\end{algorithm} 

Since it is difficult to use the common feasibility check algorithm to find a nonconvex feasible region, Algorithm \ref{Algorithm1} follows the core logic of the Douglas-Rachford scheme, which represents a method for solving the nonconvex feasibility check problem \cite{Body_2}.
We set different feasible regions $\mathcal{D}$ and $\mathcal{P}$, which satisfy different constraints to $\mathbf{x}_{1}[n]$ and $\mathbf{x}_{2}[n]$ and then minimize the distance between $\mathbf{x}_{1}[n]$ and $\mathbf{x}_{2}[n]$. \textcolor{black}{The optimization problem in Algorithm \ref{Algorithm1} can be sequentially solved by updating the feasible regions $\mathcal{D}$ and $\mathcal{P}$ for each iteration via successive convex approximation.}
\textcolor{black}{
	Through Algorithm \ref{Algorithm1}, we find the feasible set where the distance between $\mathbf{x}_{1}[n]$ and $\mathbf{x}_{2}[n]$ converges to $\epsilon$ (i.e., both feasible sets of $\mathbf{x}_{1}[n]$ and $\mathbf{x}_{2}[n]$ are satisfied).
}
\textcolor{black}{
	Note that, in the case of the distance between $\mathbf{x}_{1}[n]$ and $\mathbf{x}_{2}[n]$ greater than a certain tolerance value, the set is determined to be infeasible.
	It means that the feasible set for the original problem in \eqref{P2} does not exist under these constraints.
	In this case, we need to change the parameters (e.g., increasing the value of $N_{\mathrm{service}}$ or decreasing the value of $R_{\mathrm{th}}$).
}

\textcolor{black}{Specifically, plugging the constraints in  \eqref{P2} into Algorithm \ref{Algorithm1}, we can then write the problem to find initial value of \eqref{P2} as}

\begin{subequations}
	\begin{eqnarray}
	\!\!\!\!\!\!\!&\!\!\!\!\!\!	\displaystyle  \min_{ \scriptsize \begin{array}{c}
		\mathcal{Q}, \mathcal{V}_{1},\\ \mathcal{A}, \mathcal{V}_{2}
		\end{array} } \!&\!\! {\sum_{N_{\mathrm{service}}} \left( \| \mathbf{v}_{1}[n]-\mathbf{v}_{2}[n] \|^{2} \right) } \label{P2_Feasibility} \\
	\!\!\!\!\!\!\!&\!\!\!\!\!\! \textrm{s.t} \!&\! [\mathbf{q}[n], \mathbf{v}_{1}[n], \mathbf{a}[n]]^{T}\in\mathcal{D}, \ \mathbf{v}_{2}[n]^{T}\in\mathcal{P}, \\  
	\!\!\!\!\!\!\!&\!\!\!\!\!\! \!&\! (n=1,\cdots,N_{\mathrm{service}}). \nonumber 
	\end{eqnarray}
\end{subequations}
where $\mathcal{D}\!=\!\left\lbrace \mathbf{x} \in \mathbb{R}^{6\times N_{\mathrm{service}}} \ | \ \eqref{C_v&a}\!-\!\eqref{C_Vmax&Amax}, \eqref{C1_P2}\right\}$, $\mathcal{P}\!=\!\left\lbrace \mathbf{x}\in \mathbb{R}^{2\times N_{\mathrm{service}}} \ | \ \eqref{C_Vmin} \right\rbrace$, $\mathcal{V}_{1} = \{ \mathbf{v}_{1}[n], \forall n \}$, and $\mathcal{V}_{2} = \{ \mathbf{v}_{2}[n], \forall n \}$.
\textcolor{black}{
	The feasible set resulting from Algorithm \ref{Algorithm1} with problem \eqref{P2_Feasibility} will be the required initial values, $\mathbf{v}^{\star}_{\mathrm{service}}[n], n \in N_{\mathrm{service}}$, which are the feasible set of \eqref{P2}.
}
Accordingly, obtained initial values can be used for initializing the problem of \eqref{P2}.

As a final remark, we summarize a detailed procedure for flight time maximization in Algorithm \ref{Algorithm2}.
% 전반적으로 bisection method를 활용하면서, 이 알고리즘에서는 entry 와 exit 과정에서의 energy를 최소화한 결과와 service region에서의 energy를 최소화한 결과가 주어진 energy의 값을 만족하는지를 확인하면서, 전체적인 flight time을 최대화한다.
In this algorithm, we maximize the flight time ensuring that the result of minimizing energy in the entry and exit processes and the result of minimizing energy in the service region meet the given energy.
Note that the flight time is updated until convergence, leveraging the bisection method.

\begin{algorithm}[]
	\caption{Flight time maximization.} \label{Algorithm2}
	\begin{algorithmic}[1]
		\State Initialize $N_{\mathrm{total}}$, upper/lower bounds $\overline{N}_{\mathrm{total}}$/$\underline{N}_{\mathrm{total}}$. 
		\State Find $E_{\mathrm{entry}}$ and $N_{\mathrm{entry}}$ with Algorithm \ref{Algorithm0}. % 
		\State Find $E_{\mathrm{exit}}$ and $N_{\mathrm{exit}}$ with Algorithm \ref{Algorithm0}. % 
		%\State \emph{(Service-time maximization in $d(R_{\mathrm{th}})$, $N_{\mathrm{service}}$)}
		\While {$\overline{N}_{\mathrm{total}}-\underline{N}_{\mathrm{total}}>\epsilon$$^\dag$} 
		\State $N_{\mathrm{service}}=N_{\mathrm{total}}-(N_{\mathrm{entry}}+N_{\mathrm{exit}})$
		\State Find initial values of velocity $\mathbf{v}_{\mathrm{service}}[n], \forall n$ with Algorithm \ref{Algorithm1}.
		\State Solve \eqref{P2} with $\mathbf{v}_{\mathrm{service}}[n], \forall n$. 
		\If {$\sum_{N_{\mathrm{total}}} P(\mathbf{v}[n],\|\mathbf{v}[n]\|,\mathbf{a}[n]) \leq \dfrac{E_{\mathrm{total}}}{\delta_t}$$^\ddag$ \vspace{0.025in} \\  \ \ \ \ } {$\underline{N}_{\mathrm{total}}=N_{\mathrm{total}}$.} 
		\Else \vspace{0.025in} \\ \ \ \ \ \ \textbf{then} \ $\overline{N}_{\mathrm{total}}=N_{\mathrm{total}}$.
		\EndIf 
		\State Update $N_{\mathrm{total}}=\dfrac{\overline{N}_{\mathrm{total}}+ \underline{N}_{\mathrm{total}}}{2}$. 
		\EndWhile
		\State \textbf{end} 
	\end{algorithmic}
	\hrule
	$^\dag$ \footnotesize $\epsilon$ is the tolerance values for algorithm's termination point.\\ 
	$^\ddag$ $E_{\mathrm{total}}=E_{\mathrm{entry}}+E_{\mathrm{service}}+E_{\mathrm{exit}}$, where $E_{\mathrm{service}}$ is the consumption energy of UAV during the time step $N_{\mathrm{service}}$.
\end{algorithm}  
%$

%%%%%%%%%%%%%%%%%%%%%%%%%%%%%%%%%%%%%%%%%%%%%%%%%%%%%%%%%%%%%%%%%%%%%%%%%%%%%%%%%%%%%%%%%%%%%%%%%%%%%%%%%
\subsubsection{Energy Efficiency Maximization and Operation-Time Minimization}
The preceding flight time maximization problem can be employed as a framework that can be extended to other applications.
This framework can be applied specifically to other optimization problems, by varying the objective function and constraints related to applications where UAV-mounted FSOC is utilized.
Namely, we can find other trajectories that optimize other objective functions.
To show the applicability of the framework to several applications of UAV-assisted wireless communications, we introduce two other applicable problems, that is, energy efficiency maximization and operation-time minimization.
First, for energy efficiency maximization, we design a route that makes UAV communications operate efficiently. 
Here, we establish and maximize an energy efficiency model through both a fixed-wing UAV energy consumption model and an FSOC data rate model.
The energy efficiency maximized path can be useful in applications such as IoT services \cite{Intro_5} and cognitive radio systems \cite{Body_7}.
Second, for the operation-time minimization, we can provide a path for the UAV-mounted FSOC such that the requested amount of data is transmitted within the minimum amount of time. 
In order to solve the operation time minimization problem, we leverage rate maximization.
The rate maximized path (i.e., operation-time minimized path) can then be applied in offloading hot-spot situations \cite{Intro_17}.
We show detailed formulas and solutions to the application-oriented optimizations in Appendix.

%%%%%%%%%%%%%%%%%%%%%%%%%%%%%%%%%%%%%%%%%%%%%%%%%%%%%%%%%%%%%%%%%%%%%%%%%%%%%%%%%%%%%%%%%%%%%%%%%%%%%%%%%
\subsubsection{Complexity Analysis for Algorithm \ref{Algorithm2}}
The computational complexity of Algorithm \ref{Algorithm2} lies mainly in Step 6 and Step 7, such that Algorithm \ref{Algorithm1} and \eqref{P2} can be addressed using sequential convex optimization.
Here, we can use already existing optimization methods, such as the interior point method \cite{Body_3}, to solve the problem. 

%\textbf{Derivation of low-complexity scheme complexity :}
Through the interior point method, the computational complexity for Algorithm \ref{Algorithm2} can be given by 

\begin{equation}
\sum^{K_{3}}_{m=1}{O\left( K_{1}(9 N_{m})^3 +  K_{2}(7 N_{m})^3 \right)},  \\
\end{equation}
where the left-hand side and the right-hand side of the Big O notation indicate the complexity of both Algorithm \ref{Algorithm1} and \eqref{P2}, respectively. 
Note that $K_1$ and $K_2$ are the number of iterations used for Step 6 and Step 7.  

In a worst-case scenario, we denote $N_{m}$ as the size of the time slots in each iteration $m$ of the bisection method, which can be derived as 

\begin{equation}
N_{m} = \dfrac{\left(2^{m}-1\right)\overline{N}+\underline{N}}{2^{m}} = \overline{N}-D\biggl(\frac{1}{2}\biggr)^{m},	
\end{equation}
where $D=\overline{N}-\underline{N}$ denotes the distance between the upper bound and the lower bound for each time slot.

According to the above equation, $N_{m}^{3}$, the computational complexity of the interior point method can then lead to 

\begin{eqnarray}
\sum^{K_{3}}_{m=1}{N_{m}^{3}} \!\!&\!\!=\!\!&\!\! \overline{N}^3 K_{3} - 3\overline{N}^2 D \biggl(1-\biggl(\frac{1}{2}\biggr)^{K_{3}}\biggr) \nonumber \\
\!\!& &\!\! + \overline{N} D^2 \biggl(1\!-\!\biggl(\frac{1}{4}\biggr)^{K_{3}}\biggr) 
- \frac{1}{7} D^3 \biggl(1\!-\!\biggl(\frac{1}{8}\biggr)^{K_{3}}\biggr), \nonumber \\
\end{eqnarray}
where $K_{3}=\mathrm{floor}\{ \mathrm{log}_{2}{\frac{\overline{N}-\underline{N}}{\varepsilon}} \}$ and $\varepsilon$ denotes the termination point size.
As a result, the computational complexity for Algorithm \ref{Algorithm2} is therefore given by 

\begin{eqnarray}
\!\!\!\!&\!\!\!\! \!\!\!&\!\!\! O \Biggl( 
\left(K_{1}9^3+K_{2}7^3\right) 
\biggl( 
\overline{N}^3 K_{3} - 3\overline{N}^2 D \biggl( 1-\left(\frac{1}{2}\right)^{K_{3}} \biggr) \nonumber \\
\!\!\!\!&\!\!\!\! \!\!\!&\!\!\! \ \ \ + \ \overline{N} D^2 \biggl( 1-\left(\frac{1}{4}\right)^{K_{3}} \biggr) 
- \frac{1}{7} D^3 \biggl( 1-\left(\frac{1}{8}\right)^{K_{3}} \biggr) 
\biggr)  
\Biggr).  \nonumber \\
\!\!\!\!&\!\!\!\! \!\!\!&\!\!\! \label{Complexity_Algorithm3}
\end{eqnarray} 

%%%%%%%%%%%%%%%%%%%%%%%%%%%%%%%%%%%%%%%%%%%%%%%%%%%%%%%%%%%%%%%%%%%%%%%%%%%%%%%%%%%%%%%%%%%%%%%%%%%%%%%%%%%%%%%%%%%%%%%%%%%%%%%%%%%%%%%%%%%%%%%%%%%%%%%%%%%%%%%%%%%%%%%%%%%%%%%%%%
\section{Low-Complexity Scheme for Flight Time Maximization} \label{Body2}

From the computation complexity of Algorithm \ref{Algorithm2}, we can confirm that complexity increases significantly as the time slot increases. 
%where information about the trajectory is stored in a memory for each iteration.
Given this motivation, we come up with a low-complexity method based on the rotation transformation method.
Without an iterative process (e.g., using the bisection method), this low-complexity scheme can find a suboptimal solution for flight time maximization with only a small number of time slots.
The proposed low-complexity approach is developed by

\begin{subequations}
	\begin{eqnarray}
	\!\!\!\!\!\!\!&\!\!\!\!\!\! \displaystyle \min_{ \scriptsize \begin{array}{c}
		\mathcal{Q}, \mathcal{V}, \mathcal{A}
		\end{array} } \!\!&\!\! \sum_{N^{'}} P(\mathbf{v}[n],\|\mathbf{v}[n]\|,\mathbf{a}[n]) 	 \label{P1_Low} \\
	%\!\!\!\!\!\!\!&\!\!\!\!\!\! \!\!&\!\!  \\ 	
	\!\!\!\!\!\!\!&\!\!\!\!\!\! \textrm{s.t} \!\!&\!\! \eqref{C_v&a}-\eqref{C_q&v&a}, \eqref{C_Vmax&Amax}-\eqref{C_Vmin} \nonumber \\
	\!\!\!\!\!\!\!&\!\!\!\!\!\! \!\!&\!\! \| \mathbf{q}[n] \| \leq d(R_{\mathrm{th}}), \label{C_Boundary_P1_Low} \\
	\!\!\!\!\!\!\!&\!\!\!\!\!\! \!\!&\!\! \mathbf{q}[N^{'}]=\mathcal{W}(\theta)\cdot\mathbf{q}[1], \label{C_Rotation_q_P1_Low} \\
	\!\!\!\!\!\!\!&\!\!\!\!\!\! \!\!&\!\! \mathbf{v}[N^{'}]=\mathcal{W}(\theta)\cdot\mathbf{v}[1], \label{C_Rotation_v_P1_Low} \\
	\!\!\!\!\!\!\!&\!\!\!\!\!\! \!\!&\!\! (n=1,\ldots, N^{'}) \nonumber
	\end{eqnarray}
\end{subequations}	
where $N^{'}$ is the number of time slots for the low-complexity scheme, which is much smaller than $N_{\mathrm{total}}$ or $N_{\mathrm{service}}$.
In \eqref{C_Rotation_q_P1_Low} and \eqref{C_Rotation_v_P1_Low}, $\mathcal{W}(\theta)=  \left[ {\begin{array}{cc}
	\mathrm{cos(\theta)} & \mathrm{sin(\theta)} \\
	-\mathrm{sin(\theta)} & \mathrm{cos(\theta)} \\
	\end{array} } \right]$ is the rotation matrix, where $\theta$ denotes the angle of rotation.  
The boundary constraint\footnote{In the low-complexity scheme, we only consider UAVs when flying inside the rate boundary. We can, therefore, use the entry energy minimization (i.e., Algorithm \ref{Algorithm0}) in addition to \eqref{C_qI&qF} and \eqref{C_vI&vF} for drawing the entry path, if necessary.} in \eqref{C_Boundary_P1_Low} are the equivalent constraint to \eqref{C1_P2}.
Also note that \eqref{C_Rotation_q_P1_Low} and \eqref{C_Rotation_v_P1_Low} are constraints for the rotation transformation method, which allows the optimized trajectory to be drawn with the low-complexity scheme, as rotated and replicated by a given $\theta$.
Additionally, \eqref{C_Rotation_q_P1_Low} and \eqref{C_Rotation_v_P1_Low} can replicate the trajectories drawn within the rate boundary.

Since constraints \eqref{C_Boundary_P1_Low}, \eqref{C_Rotation_q_P1_Low}, and \eqref{C_Rotation_v_P1_Low} are all convex, we can solve \eqref{P1_Low} using the sequential optimization updating $\{\mathbf{v}_j[n]\}$ after using the slack variable and the first Taylor approximation, as in \eqref{P1}. 
From the results of \eqref{P1_Low} and the rotation transformation, it is possible to establish a trajectory of desired time slot size with less computation complexity such that

\begin{equation}
\mathbf{q}_{\mathrm{lc}}=\left(\mathbf{q}^{*} \mathbf{q}_{1} \ \mathbf{q}_{2} \ \cdots  \right), 
\end{equation}
where each position matrix in $\mathbf{q}_{\mathrm{lc}}$ is expressed as 

\begin{eqnarray}
\!\!&\!\! \!\!&\!\! \mathbf{q}_{1}[n] = \mathcal{W}(\theta)\cdot \mathbf{q}^{*}[n], \ (n=1, \cdots, N^{'}), \\
\!\!&\!\! \!\!&\!\! \mathbf{q}_{i+1}[n] = \mathcal{W}(\theta)\cdot \mathbf{q}_{i}[n],  \ (n=1, \cdots, N^{'}), 
\end{eqnarray}
Note that $\mathbf{q}^{*}$ is the optimal value of $\mathbf{q}=( \mathbf{q}[1] \ \mathbf{q}[2] \ \cdots \ \mathbf{q}[N^{'}] ) \in \mathbb{R}^{2\times N^{'}}$, when the result of \eqref{P1_Low} and $i$ is the number of rotations. 
Also note that the optimal trajectory for the UAV in the low-complexity scheme, $\mathbf{q}_{\mathrm{lc}}$, can be readily resized using $N^{'}$ and $\mathcal{W}(\theta)$.\footnote{In addition to $\mathbf{q}_{\mathrm{lc}}$, the optimal UAV velocity and acceleration for the low-complexity scheme, $\mathbf{v}_{\mathrm{lc}}$ and $\mathbf{a}_{\mathrm{lc}}$, are designed in the same way.}

In summary, if this low complexity scheme finds optimal values (e.g., $\mathbf{q}_{\mathrm{lc}}$, $\mathbf{v}_{\mathrm{lc}}$, and $\mathbf{a}_{\mathrm{lc}}$) with a much lower number of time slots $N ^{'}$, we can use the rotation transformation with $\mathcal{W}(\theta)$ to draw a patterned trajectory set to meet the desired time slot size.

%\textbf{Complexity Analysis for the Low-Complexity Scheme:}
\subsection{Complexity Analysis for the Low-Complexity Scheme}
Based on the computational complexity for Algorithm \ref{Algorithm2}, we can analyze the complexity for a low-complexity scheme.
Since the low-complexity scheme operates without the need to use the bisection method and the process of finding the initial values, the iteration numbers $K_{1}$ and $K_3$ can therefore be ignored.
The size of the time slots can then be fixed at $N^{'}$ (i.e., $K_{1}=0$, $K_{3}=1$ and $N_{m}=N^{'}$), where we can derive the complexity for the low-complexity scheme as 

\begin{equation}
O \left( K_{2} \left( 7 N^{'} \right)^{3} \right). \label{comp_low}
\end{equation}

Compared to the complexity of Algorithm \ref{Algorithm2}, \eqref{comp_low} shows that the low-complexity scheme works in less computational effort. 

%%%%%%%%%%%%%%%%%%%%%%%%%%%%%%%%%%%%%%%%%%%%%%%%%%%%%%%%%%%%%%%%%%%%%%%%%%%%%%%%%%%%%%%%%%%%%%%%%%%%%%%%%%%%%%%%%%%%%%%%%%%%%%%%%%%%%%%%%%%%%%%%%%%%%%%%%%%%%%%%%%%%%%%%%%%%%%%%%%
%%%%%% Numerical Result %%%%%%
%%%%%%%%%%%%%%%%%%%%%%%%%%%%%%%%%%%%%%%%%%%%%%%%%%%%%%%%%%%%%%%%%%%%%%%%%%%%%%%%%%%%%%%%%%%%%%%%%%%%%%%%%%%%%%%%%%%%%%%%%%%%%%%%%%%%%%%%%%%%%%%%%%%%%%%%%%%%%%%%%%%%%%%%%%%%%%%%%%
\section{Numerical Results} \label{numerical result}

\begin{figure}[]
	%    \vspace{-0.4em}
	\centering
	\includegraphics[width=3.5in]{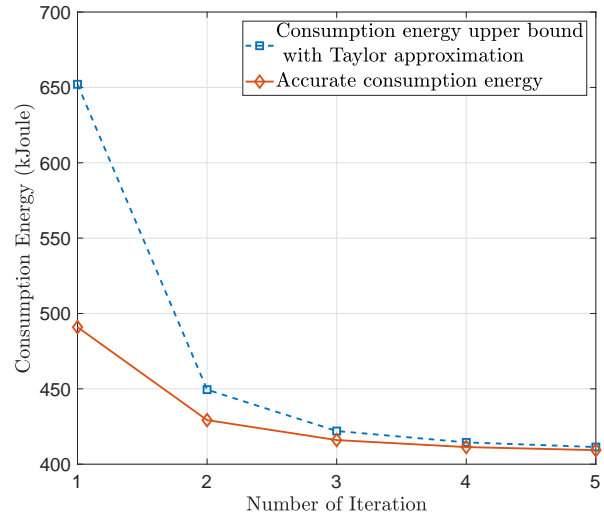}
	%    \vspace{-0.5em}  
	\caption{Convergence plot of Algorithm \ref{Algorithm2}.}
	\label{fig:Convergence}
	%   \vspace{-1em}
\end{figure}

\begin{figure}[]
	%    \vspace{-0.4em}
	\centering
	\includegraphics[width=3.5in]{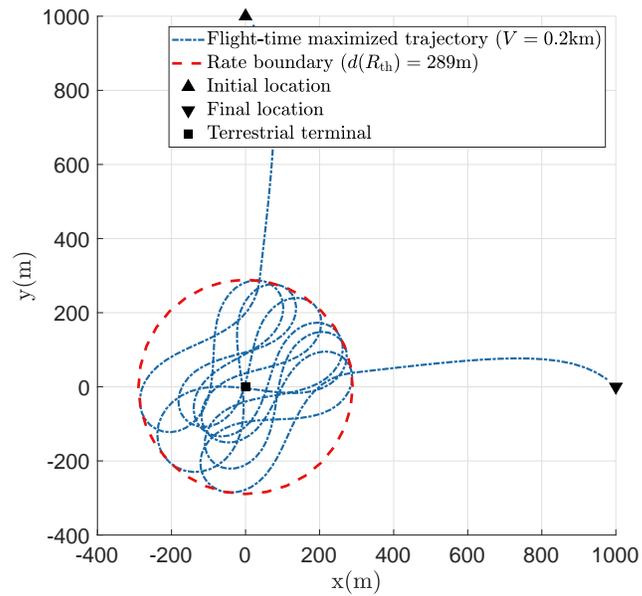}
	%    \vspace{-0.5em}  
	\caption{Maximized flight time trajectory under heavy fog ($V=0.2$ km).}
	\label{fig:dRth300}
	%   \vspace{-1em}
\end{figure}

\begin{figure}[]
	%    \vspace{-1em}
	\centering
	\includegraphics[width=3.5in]{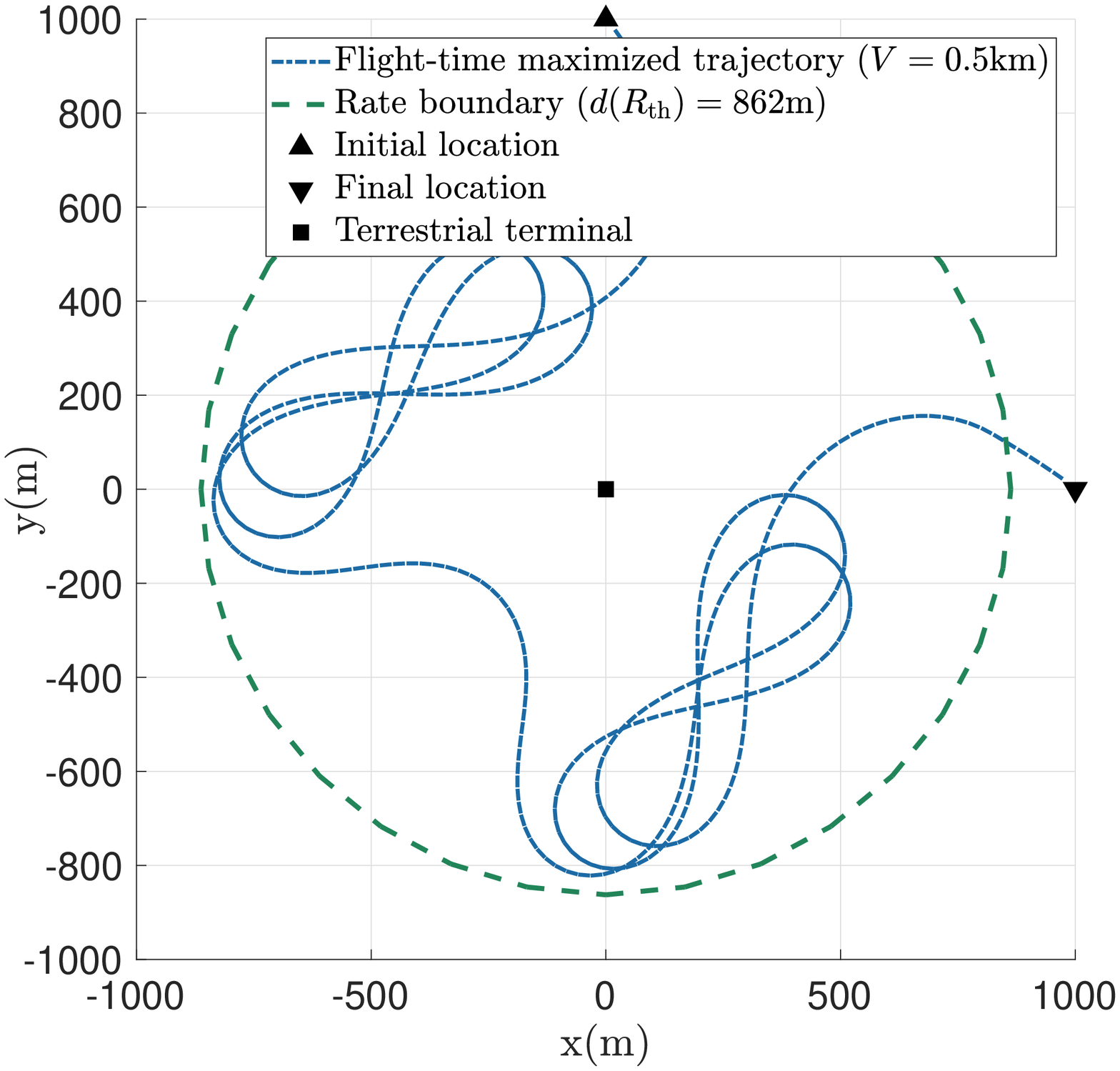}
	%    \vspace{-0.5em}  
	\caption{Maximized flight time trajectory under moderate fog ($V=0.5$ km).}
	\label{fig:dRth800}
	%   \vspace{-1em}
\end{figure}

% Set-Up
In this section, we present the numerical results of each algorithm proposed in Sections \ref{Body1} and \ref{Body2}. 
\textcolor{Black}{In the same way as in \cite{System_1}, the altitudes of the UAV and the terrestrial terminal are assumed to be constant at $H_{U}=110$ [m] and $H_{T}=10	$ [m], respectively, and we set $c_1=9.26\times10^{-4}$ [kg/m] and $c_2=2250$ [$\mathrm{kg}\cdot\mathrm{m}^3/\mathrm{s}^4$].}
We also consider the bandwidth for FSOC as $B_{\mathrm{FSO}}=1$ [MHz], with the initial location $\mathbf{q}_{\mathrm{I}}=[0,1000]^{T}$, final location $\mathbf{q}_{\mathrm{F}}=[1000,0]^{T}$, initial/final velocity $\mathbf{v}_{\mathrm{I}}=\mathbf{v}_{\mathrm{F}}=30\left( \mathbf{q}_{\mathrm{F}}-\mathbf{q}_{\mathrm{I}}\right)/\| \mathbf{q}_{\mathrm{F}}-\mathbf{q}_{\mathrm{I}}\|$, minimum velocity $V_{\mathrm{min}}=3$ [m/s], maximum velocity $V_{\mathrm{max}}=100$ [m/s], maximum acceleration $A_{\mathrm{max}}=5$ [m/$\mathrm{s}^2$],  and time-step size $\delta_t=1$ [sec]. 
As an numerical example, we set the ASNR as $\gamma=30 \ \mathrm{[dB]}$ corresponding to $\alpha=\frac{1}{4}$ as in \cite{Num_1} and also consider the minimum data rate requirement arbitrarily chosen at $R_{\mathrm{th}} = 7.94$ [Mbps].
The simulation results of this paper are obtained using the CVX.

% Convergence 설명
Fig. \ref{fig:Convergence} represents the convergence of Algorithm \ref{Algorithm2}, under this set up. 
In particular, this figure shows the convergence of the sequential optimization, used in solving \eqref{P1} and \eqref{P2}, where the terminating threshold is set as $\epsilon = 0.1 \%$.
Note that the sequential optimization has been proven to converge to at least one local optimal point \cite{System_1}.
\textcolor{Black}{Fig. \ref{fig:Convergence} consists of two curves: the upper bound of energy consumption model assuming $\mathbf{a}^T[n] \cdot \mathbf{v}[n] = 0$ which corresponds to the optimal value of \eqref{P2}, and the exact energy consumption model based on \eqref{Energy}. 
	It can be found that these two curves converge after $4$ iterations.}

\begin{figure}[]
	%    \vspace{-1em}
	\centering
	\includegraphics[width=3.5in]{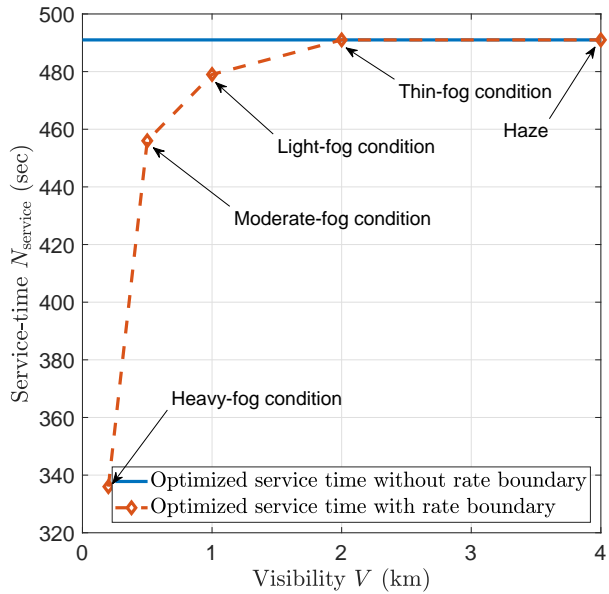}
	%    \vspace{-0.5em}  
	\caption{Maximized service time over atmospheric conditions.}
	\label{fig:NsoverV}
	%   \vspace{-1em}
\end{figure}

\begin{table*}[]
	\centering
	\caption{Performance analysis of different trajectories for the UAV.}
	\label{Table1}
	\begin{tabular}{l|c|c|c|}
		\cline{2-4}
		& \begin{tabular}[c]{@{}c@{}}Spectral efficiency\\ (bps/Hz)\end{tabular} & \begin{tabular}[c]{@{}c@{}}Consumption energy for\\ $N_{\mathrm{service}}=400$ sec (kJoule)\end{tabular} & \begin{tabular}[c]{@{}c@{}}Ratio of service time and $E_{\mathrm{total}}$\\ (s/kJoule)\end{tabular} \\ \hline
		\multicolumn{1}{|l|}{Straight path}                                                                                                                 & 8.4982                                                                 & 484.4203                                                                                               & 0.8257                                                                                              \\ \hline
		\multicolumn{1}{|l|}{Circular path ($d_{cir}=862$ m)}                                                                                               & 7.9357                                                                 & 65.8696                                                                                                & 6.0726                                                                                              \\ \hline
		\multicolumn{1}{|l|}{\begin{tabular}[c]{@{}l@{}}Maximized flight time path \\ with the low-complexity scheme ($d(R_{\mathrm{th}})=862$ m)\end{tabular}} & 8.0652                                                                 & 53.3715                                                                                                & 7.4946                                                                                              \\ \hline
		\multicolumn{1}{|l|}{Maximized flight time path ($d(R_{\mathrm{th}})=862$ m)}                                                                       & 8.3991                                                                 & 45.7058                                                                                                & \textbf{8.7516}                                                                                     \\ \hline
	\end{tabular}
	\vspace{-0em}
\end{table*}

In Figs. \ref{fig:dRth300} and \ref{fig:dRth800}, we show the trajectories in accordance with the proposed flight time maximization. 
In these figures, we set the rate boundary with $d(R_{\mathrm{th}})=289$ [m] and $d(R_{\mathrm{th}})=862$ [m], for considering both the heavy fog with $V=0.2$ [km] and the moderate fog with $V=0.5$ [km], respectively.
The total energy, given as $E_{\mathrm{total}}=50$ [kJoule], is assumed for the flight time maximization. 
Figs. \ref{fig:dRth300} and \ref{fig:dRth800} also show how the flight time can be maximized depending on the values of $d(R_{\mathrm{th}})$ set by $\gamma$, $V$, and $R_{\mathrm{th}}$. 
These results show that the time-efficient trajectory drawn by the proposed algorithm is roughly symmetrical.

%%%% NsoverV 설명
Fig. \ref{fig:NsoverV} shows the maximized service time results for different atmospheric conditions (i.e., visibility). 
% E_total = 50 [kJoule] 을 가정하였다.
Since the service boundary that satisfies service requirements changes depending on the atmospheric condition, therefore, we obtain different optimized trajectory and service time results, according to the visibility range values corresponding to weather conditions of \cite{System_6}.
In this figure, when moderate-fog condition ($V=0.8$ km), UAV-mounted FSOC can fly and provide service to terrestrial terminal $35.71 \%$ longer than heavy-fog condition ($V=0.2$ km).
It is found that, if $V$ is greater than $2$ km (thin-fog condition), optimized service time result approaches the upper bound, i.e., optimized service time result without rate boundary constraint.

\begin{figure}[t]  
	%    \vspace{-0.5em}
	\centering
	\includegraphics[width=3.5in]{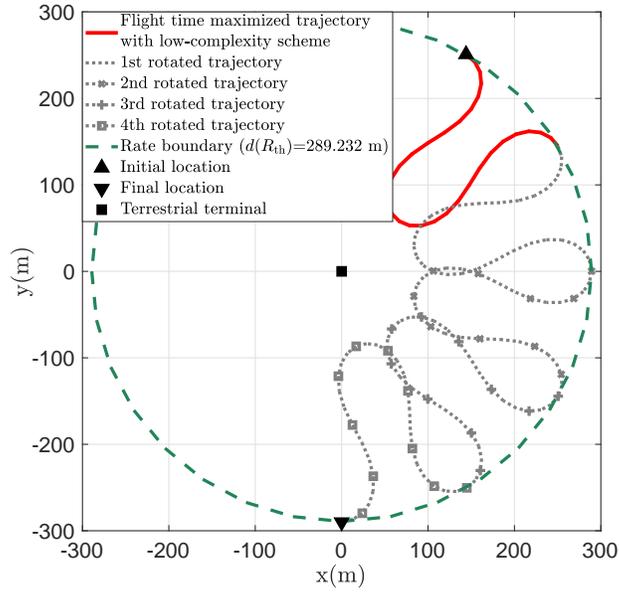}
	%    \vspace{-0.5em}  
	\caption{Maximized flight time trajectory with the low-complexity scheme under moderate fog ($V=0.5$ km).}
	\label{fig:dRth300_Low}
	%    \vspace{-1em}
\end{figure}

\begin{figure}[]
	%    \vspace{-1em}
	\centering
	\includegraphics[width=3.5in]{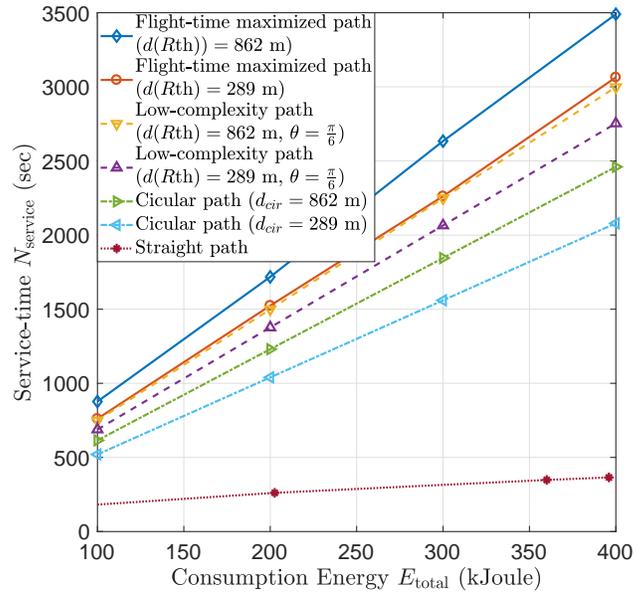}
	%    \vspace{-0.5em}  
	\caption{Relationship between service time and energy consumption for various schemes.}
	\label{fig:comparison}
	%    \vspace{-1em}
\end{figure}

%%% low complexity 방식 설명(parameter, insight)
In Fig. \ref{fig:dRth300_Low}, we show the trajectory in accordance with the proposed low-complexity scheme. 
If we consider $V=0.5$ [km], then the rate boundary is drawn as $d(R_{\mathrm{th}})=289$ [m].
This figure also shows a path for $N=150$ [sec], using only $N^{'}=30$ [sec] and $\theta=\frac{\pi}{6}$ [rad].
Bear in mind that we obtained $\theta$ numerically for the simulation.
Using this scheme, the trajectory for any desired time slot $N$ can be drawn with relatively small complexity, as shown in this figure.

Table \ref{Table1} compares the results of the UAV flying with different schemes in the same communication and atmospheric conditions. 
Here, the flight radius $d(R_{\mathrm{th}})=862$ [m] is drawn with the moderate fog condition, $V=0.5$ [km]. 
The straight path represents the path from the initial point to the terrestrial terminal and then toward the final point, at constant speed.
The circular path presents the path moving at constant speed, in the form of a circle with a radius of $d_{\mathrm{cir}}=862$ [m]. 
The proposed maximized flight time for the low-complexity scheme is drawn with a trajectory of $N^{'}=30$ [sec] and $\theta=\frac{\pi}{6}$ [rad].
The proposed maximized flight time path represents the optimized trajectory if Algorithm \ref{Algorithm2} is used.
The results of Table \ref{Table1} are based on $N_{\mathrm{service}}=400$ [sec], to consider sufficient operation time of each trajectory.
As shown in Table \ref{Table1}, the proposed maximized flight time path results in a gain of approximately 44.12\%, compared to the circular path, for the service time and consumption energy ratio.
Furthermore, the proposed maximized flight time path with the low-complexity scheme presents a gain of approximately 23.42\% over the circular path.
It can be seen that the time efficiency validates the low-complexity scheme, with reasonable performance compared to Algorithm \ref{Algorithm2}.

Following Table \ref{Table1}, Fig. \ref{fig:comparison} compares the service time versus the energy consumption for the maximized flight time path, the maximized flight time path with the low-complexity scheme, the circular path, and the straight path, where the latter of the two are conventional schemes. 
It can be seen that the proposed maximized flight time paths provide significantly more service time than conventional trajectories do.
Furthermore, Fig. \ref{fig:comparison} shows how the time efficiency ratio of the UAV changes in accordance with the given $d(R_{\mathrm{th}})$, whose value is determined using $V$ and $\gamma$. 
Intuitively, we are able to confirm that better atmospheric conditions result in a larger service range (i.e., service boundary) and better time efficiency.

%%%%%%%%%%%%%%%%%%%%%%%%%%%%%%%%%%%%%%%%%%%%%%%%%%%%%%%%%%%%%%%%%%%%%%%%%%%%%%%%%%%%%%%%%%%%%%%%%%%%%%%%%%%%%%%%%%%%%%%%%%%%%%%%%%%%%%%%%%%%%%%%%%%%%%%%%%%%%%%%%%%%%%%%%%%%%%%%%%
\section{Conclusion} \label{conclusion}
In this study, we optimized the flight time for a fixed-wing UAV-mounted FSOC. 
Based on previously proposed channels and rate models for FSOC, we investigated the flight-time-efficient trajectories over different atmospheric conditions (e.g., moderate-fog and heavy-fog conditions).
We also proposed a low-complexity method based on the rotation transformation technique and presented a complexity analysis for each scheme.
We further derived the energy efficiency maximization and operation-time minimization as extension cases of this framework.
In order to address such problems, we successively implemented several optimization methods (i.e., bisection, sequential programming, and feasibility check algorithm). 
Here, the simulation results showed the maximized flight time trajectory drawn under different atmospheric conditions. 
The numerical results demonstrated that the proposed schemes provide more time-efficient paths when compared with those of conventional schemes (i.e., straight and circular paths).

%%%%%%%%%%%%%%%%%%%%%%%%%%%%%%%%%%%%%%%%%%%%%%%%%%%%%%%%%%%%%%%%%%%%%%%%%%%%%%%%%%%%%%%%%%%%%%%%%%%%%%%%%%%%%%%%%%%%%%%%%%%%%%%%%%%%%%%%%%%%%%%%%%%%%%%%%%%%%%%%%%%%%%%%%%%%%%%%%%
\appendix

%%%%%%%%%%%%%%%%%%%%%%%%%%%%%%%%%%%%%%%%%%%%%%%%%%%%%%%%%%%%%%%
\section{Application-Oriented Optimizations}  \label{Appendix A}
\subsection{Energy Efficiency Maximization}
In this appendix, we describe the optimization of the energy efficiency for the UAV with FSOC. 
In order to solve this problem, we follow the method described in \cite{System_1, App_1}, where the UAV's energy efficiency is optimized for RF communications.
Note that the energy efficiency (EE) for FSOC using \eqref{Energy} and \eqref{Rate} can be defined as $\mathrm{EE}(\mathbf{q}[n]) \triangleq \frac{\sum_{n=1}^{N} R_{\mathrm{FSO} }(\mathbf{q}[n])\cdot \delta_t}{E_T}$.
The EE maximization problem can, therefore, be formulated mathematically as follows

\begin{eqnarray}
\!\!&\!\!	 \displaystyle \max_{ \scriptsize \begin{array}{c} 
	\mathcal{Q},\mathcal{V},\mathcal{A}
	\end{array} }  \!\!&\!\! \dfrac{\sum_{n=1}^{N} R_{\mathrm{FSO}}(\mathbf{q}[n])}	{\sum_{n=1}^{N} P(\mathbf{v}[n],\|\mathbf{v}[n]\|,\mathbf{a}[n])} \label{P1_EE}\\  
\!\!&\!\!\textrm{s.t} \!\!&\!\! \eqref{C_v&a}-\eqref{C_Vmin}. \nonumber \\
\!\!&\!\!   \!\!&\!\! (n=1,\ldots,N). \nonumber 
\end{eqnarray}

To tackle the nonconvex problem \eqref{P1_EE}, we particulary follow \cite[(P2.1)]{System_1}, \cite{App_1}, for the energy model $E_T$ and the rate model $R_{\mathrm{FSO}}(\mathbf{q}[n])$ which are the the denominator of EE and the numerator of EE, respectively.
Thus, we can readily write the EE maximization problem in \eqref{P1_EE} as

\begin{subequations}
	\begin{eqnarray}	
	\!\!&\!\!	 \displaystyle \max_{ \scriptsize \begin{array}{c}
		\mathcal{Q},\mathcal{V}\\
		\mathcal{A},\mathcal{\tau}
		\end{array} } \!\!\!&\!\!\! \dfrac{\sum_{n=1}^{N} R^{'}_{\mathrm{FSO}}(\mathbf{q}[n])}			{\sum_{n=1}^{N} P(\mathbf{v}[n],\tau[n],\mathbf{a}[n]) } \label{P2_EE} \\
	& \textrm{s.t} & \eqref{C_v&a}-\eqref{C_Vmax&Amax}, \nonumber \\
	& & V_{\min}\leq \mathbf{\tau}[n], \ \mathbf{\tau}[n]^{2} \leq \phi(\mathbf{v}[n]), \label{C1_P2_EE}   \\
	&   & (n=1,\ldots,N).  \nonumber
	\end{eqnarray}
\end{subequations}
Note that, using high-SNR approximation\footnote{Even in low visibility conditions which is the worst condition (e.g., heavy-fog condition), $k_1 \cdot e^{ -k_2 \cdot d[n] }\gg1$ is assumed within the flight range of UAV.}, the lower bound of discrete-time rate model for FSOC in [bps] can be expressed as $R^{'}_{\mathrm{FSO}}(\mathbf{q}[n])=\frac{B_{\mathrm{FSO}}}{2log2} \Big( \log(k_1)-k_2\sqrt{(H_{U}-H_{T})^2+||\mathbf{q}[n]||^2} \Big)$.

%Even in low visibility conditions which is the worst condition (e.g., heavy-fog condition), it can be found that the part of $R_{\mathrm{FSO}}[n]$ satisfies $k_1 \cdot e^{ -k_2 \cdot d[n] }\gg1$ within the flight range of UAV.

The optimization for \eqref{P2_EE} is formulated as a quadratic fractional optimization problem concerning both numerator and denominator.
Given the fractional functions with a concave numerator and a convex denominator in \eqref{P2_EE}, we can, therefore, solve  the energy efficiency maximization using the sequential convex optimization technique by iteratively updating the local point $\mathbf{v}_j[n]$.
Specifically, \eqref{P2_EE} can be addressed efficiently  via the standard Dinkelbach's algorithm for fractional programming as in \cite{Body_4}.

\subsection{Operation-Time Minimization}
In this case, we deal with the operation-time minimization that presents a trajectory for the UAV to provide the requested total data amount $\widehat{R}$ in the shortest amount of time.
In a situation where it is necessary to transfer large amounts of data within a short time frame (e.g., offloading hot-spot), the operation-time-minimized path can be utilized.
Note that the operation-time ($N_{\mathrm{operation}}$) is the total time that the data is transmitted while the UAV communication operates from the initial position to the final position.
Consequently, operation-time minimization can be briefly described as follows

\begin{subequations}
	\begin{eqnarray}
	\!\!&\!\!	 \displaystyle \min_{ \scriptsize \begin{array}{c}
		\mathcal{Q}, \mathcal{V}, \mathcal{A}
		\end{array} } \!\!&\!\! \ N_{\mathrm{operation}}	\label{P1_Operation} \\ 	
	\!\!&\!\! \textrm{s.t} \!\!&\!\! \eqref{C_v&a}-\eqref{C_Vmin}, \nonumber \\
	\!\!&\!\! \!\!&\!\! \sum_{N_{\mathrm{operation}}}R_{\mathrm{FSO}}(\mathbf{q}[n]) \geq \widehat{R}, \label{C1_P1_Operation} \\
	\!\!&\!\!   \!\!&\!\! (n=1,\ldots,N_{\mathrm{operation}}). \nonumber 
	\end{eqnarray}		
\end{subequations}

For a given $N_{\mathrm{operation}}$, the operation-time minimization problem \eqref{P1_Operation} can be understood as a rate maximization problem. 
In this case, we can replace the objective function with the FSOC rate model \eqref{Rate} for a given $N_{\mathrm{operation}}$.
Thus, \eqref{P1_Operation} can be reformulated by the rate maximization, as follow

\begin{eqnarray}
\!\!&\!\!	 \displaystyle \max_{ \scriptsize \begin{array}{c} 
	\mathcal{Q},\mathcal{V}, \mathcal{A}
	\end{array} }  \!\!&\!\! \sum_{N_{\mathrm{operation}}} R^{'}_{\mathrm{FSO}}(\mathbf{q}[n]) \label{P2_Operation}\\  		
\!\!&\!\! \textrm{s.t} \!\!&\!\! \eqref{C_v&a}-\eqref{C_Vmax&Amax}, \eqref{C1_P2_EE},  \nonumber \\
\!\!&\!\!   \!\!&\!\! (n=1,\ldots,N_{\mathrm{operation}}). \nonumber 
\end{eqnarray}	
Note that we assume in \eqref{P2_Operation} that the UAV has sufficient energy required for the flight operation.

The rate maximization problem in \eqref{P2_Operation} can, therefore, be solved in the same way as in \eqref{P2_EE}.
As a result of \eqref{P2_Operation}, we can confirm that the maximum amount of data can be transmitted for a given operation time $N_{\mathrm{operation}}$.
We can also use the bisection method to find the minimum $N_{\mathrm{operation}}$ that satisfies the constraints in \eqref{C1_P1_Operation}.
In this method, each iteration performs the following steps:

\begin{enumerate}[i)]
	\item In a given $N_{\mathrm{operation}}$, check whether the result of \eqref{P2_Operation} meets the required amount of data $\widehat{R}$. 
	\item If the rate requirement constraint \eqref{C1_P1_Operation} is satisfactory, return $N_{\mathrm{operation}}$ as the minimum operation-time and stop iteration.
	\item Otherwise, update $N_{\mathrm{operation}}$. 
\end{enumerate}

% Can use something like this to put references on a page
% by themselves when using endfloat and the captionsoff option.
\ifCLASSOPTIONcaptionsoff
  \newpage
\fi

\bibliographystyle{IEEEtran} % style de la bibliographie
\bibliography{main}

% Generated by IEEEtran.bst, version: 1.14 (2015/08/26)
\begin{thebibliography}{10}
\providecommand{\url}[1]{#1}
\csname url@samestyle\endcsname
\providecommand{\newblock}{\relax}
\providecommand{\bibinfo}[2]{#2}
\providecommand{\BIBentrySTDinterwordspacing}{\spaceskip=0pt\relax}
\providecommand{\BIBentryALTinterwordstretchfactor}{4}
\providecommand{\BIBentryALTinterwordspacing}{\spaceskip=\fontdimen2\font plus
\BIBentryALTinterwordstretchfactor\fontdimen3\font minus
  \fontdimen4\font\relax}
\providecommand{\BIBforeignlanguage}[2]{{%
\expandafter\ifx\csname l@#1\endcsname\relax
\typeout{** WARNING: IEEEtran.bst: No hyphenation pattern has been}%
\typeout{** loaded for the language `#1'. Using the pattern for}%
\typeout{** the default language instead.}%
\else
\language=\csname l@#1\endcsname
\fi
#2}}
\providecommand{\BIBdecl}{\relax}
\BIBdecl

\bibitem{Intro_1}
Y.~Zeng, R.~Zhang, and T.~J. Lim, ``Wireless communications with unmanned
  aerial vehicles: Opportunities and challenges,'' \emph{IEEE Commun. Mag.},
  vol.~54, no.~5, pp. 36--42, May. 2016.

\bibitem{Intro_18}
J.~Lyu, Y.~Zeng, and R.~Zhang, ``{UAV}-aided offloading for cellular hotspot,''
  \emph{IEEE Trans. Wireless Commun.}, vol.~17, no.~6, pp. 3988--4001, Jun.
  2018.

\bibitem{Intro_9}
I.~B.-Yaliniz and H.~Yanikomeroglu, ``The new frontier in {RAN} heterogeneity:
  Multi-tier drone-cells,'' \emph{IEEE Commun. Mag.}, vol.~54, no.~11, pp.
  48--55, Nov. 2016.

\bibitem{Intro_2}
P.~Zhan, K.~Yu, and A.~L. Swindlehurst, ``Wireless relay communications with
  unmanned aerial vehicles: Performance and optimization,'' \emph{IEEE Trans.
  Aerosp. Electron. Syst.}, vol.~47, no.~3, pp. 2068--2085, Jul. 2011.

\bibitem{Intro_8}
M.~Alzenad, M.~Z. Shakir, H.~Yanikomeroglu, and M.-S. Alouini, ``{FSO}-based
  vertical backhaul/fronthaul framework for {5G+} wireless networks,''
  \emph{IEEE Commun. Mag.}, vol.~56, no.~1, pp. 218--224, Jan. 2018.

\bibitem{Intro_10}
D.~Yanjie, M.~Z. Hassan, J.~Cheng, J.~Hossain, and V.~C.~M. Leung, ``An edge
  computing empowered radio access network with {UAV}-mounted {FSO} fronthaul
  and backhaul: Key challenges and approaches,'' \emph{arXiv:1803.06381
  [cs.NI]}, pp. 1--15, Mar. 2018.

\bibitem{Intro_11}
{The Internet Research Task Force}, ``{IRTF Global Access to the Internet for
  All Research Group (GAIA)},'' \url{https://irtf.org/gaia}, accessed
  2018-11-12.

\bibitem{Intro_13}
B.~Newton, J.~Aikat, and K.~Jeffay, ``Explicit topology management for
  continental-scale airborne networks,'' in \emph{Proc. Int. Conf. on Computer
  Communication (INFOCOM)}, Atlanta, GA, USA, May 2017, pp. 72--77.

\bibitem{Intro_12}
T.~Peyronel, K.~J. Quirk, S.~C. Wang, and T.~G. Tiecke, ``Luminescent detector
  for free-space optical communication,'' \emph{Optica.}, vol.~3, no.~7, pp.
  787--792, Jul. 2016.

\bibitem{Intro_6}
\BIBentryALTinterwordspacing
{Facebook Code}, ``Building communications networks in the stratosphere.''
  {A}ccessed: 11- Sep- 2018. [Online]. Available:
  \url{https://code.fb.com/connectivity/building-communications-networks-in-the-stratosphere/}
\BIBentrySTDinterwordspacing

\bibitem{Intro_3}
Q.~Wu, Y.~Zeng, and R.~Zhang, ``Joint trajectory and communication design for
  multi-{UAV} enabled wireless networks,'' \emph{IEEE Trans. Wireless Commun.},
  vol.~17, no.~3, pp. 2109--2121, Mar. 2018.

\bibitem{Intro_4}
S.~Jeong, O.~Simeone, and J.~Kang, ``Mobile edge computing via a {UAV}-mounted
  cloudlet: Optimization of bit allocation and path planning,'' \emph{IEEE
  Trans. Veh. Technol.}, vol.~67, no.~3, pp. 2049--2062, Mar. 2018.

\bibitem{Intro_5}
M.~Mozaffari, W.~Saad, M.~Bennis, and M.~Debbah, ``Mobile unmanned aerial
  vehicles ({UAV}s) for energy-efficient internet of things communications,''
  \emph{IEEE Trans. Wireless Commun.}, vol.~16, no.~11, pp. 7574--7589, Nov.
  2017.

\bibitem{Intro_20}
R.~I. Bor-Yaliniz, A.~El-Keyi, and H.~Yanikomeroglu, ``Efficient 3{-D}
  placement of an aerial base station in next generation cellular networks,''
  in \emph{Proc. International Conf. on Commun.}, Kuala Lumpur, Malaysia, May.
  2016, pp. 1--5.

\bibitem{Intro_7}
W.~Fawaz, C.~Abou-Rjeily, and C.~Assi, ``{UAV}-aided cooperation for {FSO}
  communication systems,'' \emph{IEEE Commun. Mag.}, vol.~56, no.~1, pp.
  70--75, Jan. 2018.

\bibitem{Intro_14}
M.~D.~Yang, Q.~Wu, Y.~Zeng, and R.~Zhang, ``Energy trade-off in ground-to-{UAV}
  communication via trajectory design,'' \emph{IEEE Trans. Veh. Technol.},
  vol.~67, no.~7, pp. 6721--6726, Jul. 2018.

\bibitem{Intro_15}
M.~Thammawichai, S.~P. Baliyarasimhuni, E.~C. Kerrigan, and J.~Sousa,
  ``Optimizing communication and computation for multi-{UAV} information
  gathering applications,'' \emph{IEEE Trans. Aerosp. Electron. Syst.},
  vol.~54, no.~2, pp. 1--10, Apr. 2018.

\bibitem{Intro_16}
H.~Ghazzai, M.~B. Ghorbel, and A.~Kadri, ``Energy efficient {3D} positioning of
  micro unmanned aerial vehicles for underlay cognitive radio systems,'' in
  \emph{Proc. International Conf. on Commun.}, Paris, France, Jul. May. 2017,
  pp. 1--6.

\bibitem{Intro_19}
M.~B. M.~Mozaffari, W.~Saad and M.~Debbah, ``Wireless communication using
  unmanned aerial vehicles ({UAV}s): Optimal transport theory for hover time
  optimization,'' \emph{IEEE Trans. Wireless Commun.}, vol.~16, no.~12, pp.
  8052--8066, Dec. 2017.

\bibitem{Intro_27}
\BIBentryALTinterwordspacing
H.~Bolandhemmat, B.~Thomsen, and J.~Marriott. (2018) Energy-optimized
  trajectory planning for high altitude long endurance {(HALE)} aircraft.
  [Online]. Available:
  \url{https://research.fb.com/publications/energy-optimized-trajectory-planning-for-high-altitude-long-endurance-hale-aircraft/}
\BIBentrySTDinterwordspacing

\bibitem{Intro_28}
\BIBentryALTinterwordspacing
J.~Marriott, B.~Tezel, Z.~Liu, and N.~Stier. (2018) Trajectory optimization of
  solar-powered high-altitude long endurance aircraft. [Online]. Available:
  \url{https://research.fb.com/publications/trajectory-optimization-of-solar-powered-high-altitude-long-endurance-aircraft/}
\BIBentrySTDinterwordspacing

\bibitem{Intro_22}
H.~{Dahrouj}, A.~{Douik}, F.~{Rayal}, T.~Y. {Al-Naffouri}, and M.~{Alouini},
  ``Cost-effective hybrid {RF/FSO} backhaul solution for next generation
  wireless systems,'' \emph{IEEE Wireless Commun.}, vol.~22, no.~5, pp.
  98--104, Oct. 2015.

\bibitem{App_1}
J.-H. Lee, K.-H. Park, M.-S. Alouini, and Y.-C. Ko, ``Trajectory optimization
  of energy efficient {FSOC-UAV} with atmospheric and geometric los,'' in
  \emph{Proc. International Conf. on Ubiquitous and Future Netw. (ICUFN)},
  Prague, Czech Republic, Jul. 2018, pp. 35--37.

\bibitem{Intro_26}
\BIBentryALTinterwordspacing
{Google LLC}. (2019) Loon project. [Online]. Available: \url{https://loon.co/}
\BIBentrySTDinterwordspacing

\bibitem{Intro_23}
S.~{Chandrasekharan}, K.~{Gomez}, A.~{Al-Hourani}, S.~{Kandeepan},
  T.~{Rasheed}, L.~{Goratti}, L.~{Reynaud}, D.~{Grace}, I.~{Bucaille},
  T.~{Wirth}, and S.~{Allsopp}, ``Designing and implementing future aerial
  communication networks,'' \emph{IEEE Commun. Mag.}, vol.~54, no.~5, pp.
  26--34, May 2016.

\bibitem{Intro_25}
Y.~{Li}, N.~{Pappas}, V.~{Angelakis}, M.~{Pióro}, and D.~{Yuan},
  ``Optimization of free space optical wireless network for cellular
  backhauling,'' \emph{IEEE J. Sel. Areas Commun.}, vol.~33, no.~9, pp.
  1841--1854, Sep. 2015.

\bibitem{Intro_24}
Z.~{Gu}, J.~{Zhang}, Y.~{Ji}, L.~{Bai}, and X.~{Sun}, ``Network topology
  reconfiguration for {FSO}-based fronthaul/backhaul in {5G+} wireless
  networks,'' \emph{IEEE Access}, vol.~6, pp. 69\,426--69\,437, 2018.

\bibitem{System_1}
Y.~Zeng and R.~Zhang, ``Energy-efficient {UAV} communication with trajectory
  optimization,'' \emph{IEEE Trans. Wireless Commun.}, vol.~16, no.~6, pp.
  3747--3760, Jun. 2017.

\bibitem{System_7}
A.~A. Farid and S.~Hranilovic, ``Outage capacity optimization for free-space
  optical links with pointing errors,'' \emph{J. Lightw. Technol.}, vol.~25,
  no.~7, pp. 1702--1710, Jul. 2007.

\bibitem{System_8}
A.~Filippone, \emph{Flight Performance of Fixed and Rotary Wing
  Aircraft}.\hskip 1em plus 0.5em minus 0.4em\relax Amsterdam, Netherlands:
  Elsevier, 2006.

\bibitem{System_2}
I.-R. P.1814, ``Prediction methods required for the design of terrestrial
  free-space optical links,'' in \emph{International Telecommunication Union},
  Geneva, Switzerland, 2007, pp. 1--12.

\bibitem{System_5}
M.~A. Esmail, H.~Fathallah, and M.-S. Alouini, ``Outdoor {FSO} communications
  under fog: Attenuation modeling and performance evaluation,'' \emph{IEEE
  Photon. J.}, vol.~8, no.~4, p. 7905622, Aug. 2016.

\bibitem{System_6}
{H. Kaushal}, {V. K. Jain}, and {S. Kar}, \emph{Free Space Optical
  Communication}.\hskip 1em plus 0.5em minus 0.4em\relax India: Springer, 1st
  ed. 2017.

\bibitem{System_10}
I.~I. Kim, B.~McArthur, and E.~J. Korevaar, ``Comparison of laser beam
  propagation at 785 nm and 1550 nm in fog and haze for optical wireless
  communications,'' in \emph{Optical Wireless Communications III}, vol. 4214,
  Boston, United States, 2001.

\bibitem{System_3}
A.~Lapidoth, S.~M. Moser, and M.~A. Wigger, ``On the capacity of free-space
  optical intensity channels,'' \emph{IEEE Trans. Inf. Theory.}, vol.~55,
  no.~10, pp. 4449--4461, Oct. 2009.

\bibitem{System_4}
A.~Chaaban, J.-M. Morvan, and M.-S. Alouini, ``Free-space optical
  communications: Capacity bounds, approximations, and a new sphere-packing
  perspective,'' \emph{IEEE Trans. Commun.}, vol.~64, no.~3, pp. 1176--1191,
  Feb. 2016.

\bibitem{System_11}
W.~H. Press, S.~A. Teukolsky, W.~T. Vetterling, and B.~P. Flannery,
  \emph{Numerical Recipes: The Art of Scientific Computing}.\hskip 1em plus
  0.5em minus 0.4em\relax New York: Cambridge University Press, 3rd ed. 2007.

\bibitem{Body_6}
M.~Grant and S.~Boyd, ``{CVX}: Matlab software for disciplined convex
  programming, version 2.1,'' \url{http://cvxr.com/cvx}, Mar. 2014.

\bibitem{Body_4}
A.~Zappone, E.~Bj{\"{o}}rnson, and L.~Sanguinetti, ``Globally optimal
  energy-efficient power control and receiver design in wireless networks,''
  \emph{IEEE Trans. Signal Process.}, vol.~65, no.~11, pp. 2844--2859, Jun.
  2017.

\bibitem{Body_5}
A.~Beck, A.~Ben-Tal, and L.~Tetruashvili, ``A sequential parametric convex
  approximation method with applications to nonconvex truss topology design
  problems,'' \emph{J. Global Optim.}, vol.~47, no.~1, pp. 29--51, May. 2010.

\bibitem{Body_1}
J.~Corvellec and S.~D. Fla{\"{o}}m, ``Non-convex feasibility problems and
  proximal point methods,'' \emph{Optimization Methods and Software}, vol.~19,
  no.~1, pp. 3--14, Feb. 2004.

\bibitem{Body_2}
G.~Li and T.~K. Pong, ``Douglas-{Rachford} splitting for nonconvex optimization
  with application to nonconvex feasibility problems,'' \emph{Math. Program.},
  vol. 159, no. 1-2, pp. 371--401, Sep. 2016.

\bibitem{Body_7}
L.~Sboui, H.~Ghazzai, Z.~Rezki, and M.-S. Alouini, ``Energy-efficient power
  allocation for {UAV} cognitive radio systems,'' in \emph{Proc. Vehicular
  Technol. Conf. (VTC-Fall)}, Toronto, ON, Canada, Sep. 2017, pp. 1--5.

\bibitem{Intro_17}
M.~Mozaffari, W.~Saad, M.~Bennis, and M.~Debbah, ``Performance optimization for
  {UAV}-enabled wireless communications under flight time constraints,'' in
  \emph{Proc. IEEE Global Commun. Conf.}, Singapore, Dec. 2017, pp. 1--6.

\bibitem{Body_3}
S.~Boyd and L.~Vandenberghe, \emph{Convex Optimization}.\hskip 1em plus 0.5em
  minus 0.4em\relax Cambridge, U.K.: Cambridge Univ. Press, 2004.

\bibitem{Num_1}
H.~AlQuwaiee, I.~S. Ansari, and M.-S. Alouini, ``On the performance of
  free-space optical communication systems over double generalized gamma
  channel,'' \emph{IEEE J. Sel. Areas Commun.}, vol.~33, no.~9, pp. 1829--1840,
  Sep. 2015.

\end{thebibliography}

% that's all folks
\end{document}